\documentclass[epj]{svjour}
\usepackage{graphics}
\begin{document}
\title{Electromagnetic radiative corrections in parity-violating electron-proton scattering}

\author{J. Arvieux\inst{1} \and B. Collin\inst{1,}\thanks{\emph{Present address:} Department of Physics and Astronomy, Pennsylvania State University, PA 16802, USA}\and H. Guler \inst{1,}\thanks{\emph{Present address:} DESY, Hambourg, Germany} \and M. Morlet\inst{1} \and S. Niccolai\inst{1} \and S. Ong{\inst{1,2}} \and J. Van de Wiele\inst{1}}
%}                    % Do not remove
\institute{ Institut de Physique Nucl\'eaire, IN2P3-CNRS, Universit\'e de
Paris-Sud, 91406 Orsay Cedex, France \and Picardie Jules Verne University, F-80000 Amiens, France}
\date{Received: date / Revised version: date}
\abstract{
QED radiative corrections have been calculated for leptonic and hadronic
variables in parity-violating elastic {\it ep} scattering. For the first time,
the calculation of the asymmetry in the elastic radiative tail is performed
without the peaking-approximation assumption in hadronic variables
configuration. A comparison with the PV-A4 data validates our approach.
This method has been also used to evaluate the radiative
corrections to the parity-violating asymmetry measured in the G0 
experiment. The results obtained are here presented. 
}
\def\gee{$G_E$}
\def\gmm{$G_M$}
\def\g0{$G^0$}
\def\pion{$\pi ^+$}
\def\q2{$Q^2$}
\def\z0{$Z^0$}
\def\gam{$\gamma \hspace{0mm}$ }

\def\Lvarepsilon{\mbox{$ \varepsilon $ } \! \! }

\newcommand{\qslash}
           {\mbox{$ q \hspace{-1.7mm} / \hspace{1.2mm} $}}
	   
\newcommand{\pslash}
           {\mbox{$ p \hspace{-1.5mm} / \hspace{1mm} $}}

\newcommand{\ppslash}
           {\mbox{$ p' \hspace{-2.6mm} / \hspace{2.1mm} $}}

\setlength{\textheight}{22.5cm}
\def\gam{$\gamma$ }
\def\gammag  { \mbox{\boldmath$\gamma$} }
\def\sigmag  { \mbox{\boldmath$\sigma$} }
\newcommand{\pnslash}
           {\mbox{$ P_{_N}  \hspace{-4.2mm} / \hspace{2.8mm} $}}
\newcommand{\funx}
           {\mbox{$F _{_{ \! 1}} ^{^{x}}$}}

\newcommand{\fdeuxx}
           {\mbox{$F _{_{ \! 2}}^{^{x}} $}}

\newcommand{\ftunx}
           {\mbox{$\widetilde F _{_{ \! 1}} ^{^{x}}$}}

\newcommand{\ftdeuxx}
           {\mbox{$\widetilde F _{_{ \! 2}}^{^{x}} $}}
\newcommand{\gtax}
           {\mbox{$\widetilde G _{_{ \! A}} ^{^x} $}}
\newcommand{\gtpx}
           {\mbox{$\widetilde G _{_{ \! P}} ^{^x} $}}

\newcommand{\gex}{\mbox{$G_{_{\! E}}^{^{x}}$}}
\newcommand{\gmx}{\mbox{$G_{_{\! M}}^{^{x}}$}}
\newcommand{\gtex}
           {\mbox{$\widetilde G _{_{ \! E}} ^{^x} $}}
\newcommand{\gtmx}
           {\mbox{$\widetilde G _{_{ \! M}} ^{^x} $}}

\newcommand{\funp}
           {\mbox{$F _{_{ \! 1}} ^{^{p}}$}}

\newcommand{\fdeuxp}
           {\mbox{$F _{_{ \! 2}}^{^{p}} $}}

\newcommand{\funn}
           {\mbox{$F _{_{ \! 1}} ^{^{n}}$}}

\newcommand{\fdeuxn}
           {\mbox{$F _{_{ \! 2}}^{^{n}} $}}

\newcommand{\geu}
           {\mbox{$G _{_{E}} ^{^{ {\scriptstyle u } }} $}}
\newcommand{\gmu}
           {\mbox{$G _{_{M}} ^{^{ {\scriptstyle u } }} $}}

\newcommand{\ged}
           {\mbox{$G _{_{E}} ^{^{ {\scriptstyle d } }} $}}
\newcommand{\gmd}
           {\mbox{$G _{_{M}} ^{^{ {\scriptstyle d } }} $}}
	   
\newcommand{\getrois}
           {\mbox{$G _{_{E}} ^{^{( 3 )}} $}}
\newcommand{\gehuit}
           {\mbox{$G _{_{E}} ^{^{( 8 )}} $}}

\newcommand{\gmtrois}
           {\mbox{$G _{_{M}} ^{^{( 3 )}} $}}
\newcommand{\gmhuit}
           {\mbox{$G _{_{M}} ^{^{( 8 )}} $}}
	   
\newcommand{\ggatequn}
           {\mbox{$   G _{_{ \! A}} ^{^(T=1)} $}}
	   
\newcommand{\ggateqzero}
           {\mbox{$   G _{_{ \! A}} ^{^(T=0)} $}}

\newcommand{\atroist}
           {\mbox{$ \tilde{a}_{_{\scriptstyle 3}} $}}
\newcommand{\ahuitt}
           {\mbox{$ \tilde{a}_{_{\scriptstyle 8}} $}}

\newcommand{\ropeq}
           {\mbox{$ {\rho}^{\prime}_{\! eq} $}} 
\newcommand{\roeq}
           {\mbox{$ {\rho}_{\! eq} $}} 

\newcommand{\hkappapeq}
           {\mbox{$  \hat{\kappa}^{\prime}_{\! eq}   $}}
\newcommand{\hkappaeq}
           {\mbox{$  \hat{\kappa}_{\! eq}   $}}

\newcommand{\hszdeux}
           {\mbox{$  \hat{s}^{2}_{\! Z}   $}}

\newcommand{\Aslash}
           {\mbox{$ A \hspace{-0.2cm} / \hspace{0.1cm} $}}
\newcommand{\calu}
           {\mbox{$ {\scriptstyle \cal U }$}}
\newcommand{\calubar}
           {\mbox{${\scriptstyle \overline {\cal U }} $}}

\newcommand{\champe}
           {\mbox{$ \psi _{e}$}}
\newcommand{\champebar}
           {\mbox{${\bar \psi}_{e}$}}

\newcommand{\champf}
           {\mbox{$ \psi _{f}$}}
\newcommand{\champfbar}
           {\mbox{${\bar \psi}_{f}$}}

\newcommand{\champu}
           {\mbox{$ \psi _{u}$}}
\newcommand{\champubar}
           {\mbox{${\bar \psi}_{u}$}}

\newcommand{\champd}
           {\mbox{$ \psi _{d}$}}
\newcommand{\champdbar}
           {\mbox{${\bar \psi}_{d}$}}

\newcommand{\champs}
           {\mbox{$ \psi _{s}$}}
\newcommand{\champsbar}
           {\mbox{${\bar \psi}_{s}$}}

\newcommand{\fdeuxng}
           {\mbox{$F _{_{ \! 2}} ^{^{n,\gamma}} $}}
\newcommand{\funpg}
           {\mbox{$F _{_{ \! 1}} ^{^{p,\gamma}} $}}
\newcommand{\fdeuxpg}
           {\mbox{$F _{_{ \! 2}} ^{^{p,\gamma}} $}}
\newcommand{\ftunp}
           {\mbox{$\widetilde F _{_{ \! 1}} ^{^p} $}}
\newcommand{\ftdeuxp}
           {\mbox{$\widetilde F _{_{ \! 2}} ^{^p} $}}
\newcommand{\gtafp}
           {\mbox{$\widetilde G _{_{A}} ^{^{f,p}} $}}
\newcommand{\ftpfp}
           {\mbox{$\widetilde G _{_{P}} ^{^{f,p}} $}}
\newcommand{\gtafn}
           {\mbox{$\widetilde G _{_{A}} ^{^{f,n}} $}}
\newcommand{\ftpfn}
           {\mbox{$\widetilde G _{_{P}} ^{^{f,n}} $}}
\newcommand{\ftunn} 
           {\mbox{$\widetilde F _{_{ \! 1}} ^{^n} $}}
\newcommand{\ftdeuxn} 
           {\mbox{$\widetilde F _{_{ \! 2}} ^{^n} $}}

\newcommand{\gtafx}
           {\mbox{$\widetilde G _{_{A}} ^{^{f,x}} $}}
\newcommand{\ftpfx}
           {\mbox{$\widetilde G _{_{P}} ^{^{f,x}} $}}

\newcommand{\gtaup}
           {\mbox{$\widetilde G _{_{A}} ^{^{u,p}} $}}
\newcommand{\gtadp}
           {\mbox{$\widetilde G _{_{A}} ^{^{d,p}} $}}
\newcommand{\gtasp}
           {\mbox{$\widetilde G _{_{A}} ^{^{s,p}} $}}
\newcommand{\gtaun}
           {\mbox{$\widetilde G _{_{A}} ^{^{u,n}} $}}
\newcommand{\gtadn}
           {\mbox{$\widetilde G _{_{A}} ^{^{d,n}} $}}
\newcommand{\gtasn}
           {\mbox{$\widetilde G _{_{A}} ^{^{s,n}} $}}

\newcommand{\gtaq}
           {\mbox{$           G _{_{A}} ^{^{q}} $}}
\newcommand{\gtaf}
           {\mbox{$           G _{_{A}} ^{^{f}} $}}
\newcommand{\gtau}
           {\mbox{$           G _{_{A}} ^{^{u}} $}}
\newcommand{\gtad}
           {\mbox{$           G _{_{A}} ^{^{d}} $}}
\newcommand{\gtas}
           {\mbox{$           G _{_{A}} ^{^{s}} $}}

\newcommand{\gtpup}
           {\mbox{$\widetilde G _{_{P}} ^{^{u,p}} $}}
\newcommand{\gtpdp}
           {\mbox{$\widetilde G _{_{P}} ^{^{d,p}} $}}
\newcommand{\gtpsp}
           {\mbox{$\widetilde G _{_{P}} ^{^{s,p}} $}}
\newcommand{\gtpun}
           {\mbox{$\widetilde G _{_{P}} ^{^{u,n}} $}}
\newcommand{\gtpdn}
           {\mbox{$\widetilde G _{_{P}} ^{^{d,n}} $}}
\newcommand{\gtpsn}
           {\mbox{$\widetilde G _{_{P}} ^{^{s,n}} $}}

\newcommand{\funfx}
           {\mbox{$F _{_{ \! 1}} ^{^{f,x}} $}}
\newcommand{\fdeuxfx}
           {\mbox{$F _{_{ \! 2}} ^{^{f,x}} $}}

\newcommand{\funfp}
           {\mbox{$F _{_{ \! 1}} ^{^{f,p}} $}}
\newcommand{\fdeuxfp}
           {\mbox{$F _{_{ \! 2}} ^{^{f,p}} $}}
\newcommand{\funfn}
           {\mbox{$F _{_{ \! 1}} ^{^{f,n}} $}}
\newcommand{\fdeuxfn}
           {\mbox{$F _{_{ \! 2}} ^{^{f,n}} $}}
\newcommand{\fafp}
           {\mbox{$F _{_{ \! A}} ^{^{f,p}} $}}
\newcommand{\fafn}
           {\mbox{$F _{_{ \! A}} ^{^{f,n}} $}}

\newcommand{\funup}
           {\mbox{$F _{_{ \! 1}} ^{^{u,p}} $}}
\newcommand{\fdeuxup}
           {\mbox{$F _{_{ \! 2}} ^{^{u,p}} $}}
\newcommand{\funsp}
           {\mbox{$F _{_{ \! 1}} ^{^{s,p}} $}}
\newcommand{\fdeuxsp}
           {\mbox{$F _{_{ \! 2}} ^{^{s,p}} $}}
\newcommand{\funun}
           {\mbox{$F _{_{ \! 1}} ^{^{u,n}} $}}
\newcommand{\fdeuxun}
           {\mbox{$F _{_{ \! 2}} ^{^{u,n}} $}}
\newcommand{\funsn}
           {\mbox{$F _{_{ \! 1}} ^{^{s,n}} $}}
\newcommand{\fdeuxsn}
           {\mbox{$F _{_{ \! 2}} ^{^{s,n}} $}}
\newcommand{\fundp}
           {\mbox{$F _{_{ \! 1}} ^{^{d,p}} $}}
\newcommand{\fdeuxdp}
           {\mbox{$F _{_{ \! 2}} ^{^{d,p}} $}}
\newcommand{\fundn}
           {\mbox{$F _{_{ \! 1}} ^{^{d,n}} $}}
\newcommand{\fdeuxdn}
           {\mbox{$F _{_{ \! 2}} ^{^{d,n}} $}}
\newcommand{\fasp}
           {\mbox{$F _{_{ \! A}} ^{^{s,p}} $}}
\newcommand{\fasn}
           {\mbox{$F _{_{ \! A}} ^{^{s,n}} $}}
\newcommand{\funu}
           {\mbox{$F _{_{ \! 1}} ^{^{u}} $}}
\newcommand{\fund}
           {\mbox{$F _{_{ \! 1}} ^{^{d}} $}}
\newcommand{\funs}
           {\mbox{$F _{_{ \! 1}} ^{^{s}} $}}
\newcommand{\fdeuxu}
           {\mbox{$F _{_{ \! 2}} ^{^{u}} $}}
\newcommand{\fdeuxd}
           {\mbox{$F _{_{ \! 2}} ^{^{d}} $}}
\newcommand{\fdeuxs}
           {\mbox{$F _{_{ \! 2}} ^{^{s}} $}}
\newcommand{\fas}
           {\mbox{$F _{_{ \! A}} ^{^{s}} $}}

\newcommand{\xyz}{\mbox{$G_{_{\! E}}$}}
\newcommand{\gm}{\mbox{$G_{_{\! M}}$}}

\newcommand{\gep}{\mbox{$G_{_{\! E}}^{^{p}}$}}
\newcommand{\gept}{\mbox{$\widetilde G_{_{\! E}}^{^{p}}$}}
\newcommand{\gmp}{\mbox{$G_{_{\! M}}^{^{p}}$}}
\newcommand{\gmpt}{\mbox{$\widetilde G_{_{\! M}}^{^{p}}$}}
\newcommand{\gen}{\mbox{$G_{_{\! E}}^{^{n}} $}}
\newcommand{\gent}{\mbox{$\widetilde G_{_{\! E}}^{^{n}}$}}
\newcommand{\gmn}{\mbox{$G_{_{\! M}}^{^{n}}$}}
\newcommand{\gmnt}{\mbox{$\widetilde G_{_{\! M}}^{^{n}}$}}
\newcommand{\get}{\mbox{$\widetilde G_{_{\! E}}$}}
\newcommand{\gmt}{\mbox{$\widetilde G_{_{\! M}}$}}

\newcommand{\fthreet}{\mbox{$\widetilde F_{_{\! 3}}$}}

\newcommand{\gezp}{\mbox{$G_{_{\! E}}^{^{p,Z}}$}}
\newcommand{\gmzp}{\mbox{$G_{_{\! M}}^{^{p,Z}}$}}
\newcommand{\gazp}{\mbox{$G_{_{\! A}}^{^{p,Z}}$}}

\newcommand{\gezn}{\mbox{$G_{_{\! E}}^{^{n,Z}}$}}
\newcommand{\gmzn}{\mbox{$G_{_{\! M}}^{^{n,Z}}$}}
\newcommand{\gazn}{\mbox{$G_{_{\! A}}^{^{n,Z}}$}}

\newcommand{\gtep}
           {\mbox{$\widetilde G _{_{ \! E}} ^{^p} $}}
\newcommand{\gtmp}
           {\mbox{$\widetilde G _{_{ \! M}} ^{^p} $}}
\newcommand{\gtap}
           {\mbox{$\widetilde G _{_{ \! A}} ^{^p} $}}
\newcommand{\gtpp}
           {\mbox{$\widetilde G _{_{ \! P}} ^{^p} $}}
\newcommand{\gtan}
           {\mbox{$\widetilde G _{_{ \! A}} ^{^n} $}}
\newcommand{\gtpn}
           {\mbox{$\widetilde G _{_{ \! P}} ^{^n} $}}
\newcommand{\gten}
           {\mbox{$\widetilde G _{_{ \! E}} ^{^n} $}}
\newcommand{\gtmn}
           {\mbox{$\widetilde G _{_{ \! M}} ^{^n} $}}

\newcommand{\lambdaes}
           {\mbox{$\lambda _{_{E}} ^{^{( {\scriptstyle s } )}} $}}
\newcommand{\lambdams}
           {\mbox{$\lambda _{_{E}} ^{^{( {\scriptstyle s } )}} $}}
\newcommand{\lambdaas}
           {\mbox{$\lambda _{_{A}} ^{^{( {\scriptstyle s } )}} $}}

\newcommand{\tautroisp}
           {\mbox{$\, $}}
\newcommand{\tautroism}
           {\mbox{$ - $}}

\newcommand{\xies}
           {\mbox{$\xi _{_{E}} ^{^{( {\scriptstyle s } )}} $}}
\newcommand{\xims}
           {\mbox{$\xi _{_{M}} ^{^{( {\scriptstyle s } )}} $}}
\newcommand{\xias}
           {\mbox{$\xi _{_{A}} ^{^{( {\scriptstyle s } )}} $}}	   

\newcommand{\xivp}
           {\mbox{$ \xi _{_V}^{^p} $ }}
\newcommand{\xivn}
           {\mbox{$ \xi _{_V}^{^n} $ }}

\newcommand{\xiatequn}
           {\mbox{$ \xi _{_A}^{^{T=1}}   $}}
\newcommand{\xiateqzero}
           {\mbox{$ \xi _{_A}^{^{T=0}}   $}}
\newcommand{\xiazero}
           {\mbox{$ \xi _{_A}^{^{(0)}}   $}}

\newcommand{\xivtequn}
           {\mbox{$ \xi _{_V}^{^{T=1} }   $}}
\newcommand{\xivteqzero}
           {\mbox{$ \xi _{_V}^{^{T=0} }   $}}
\newcommand{\xivzero}
           {\mbox{$ \xi _{_V}^{^{(0)} }   $}}           
\newcommand{\ges}
           {\mbox{$G _{_{E}} ^{^{ {\scriptstyle s } }} $ }}
\newcommand{\gms}
           {\mbox{$G _{_{M}} ^{^{ {\scriptstyle s } }} $ }}
\newcommand{\gas}
           {\mbox{$G _{_{A}} ^{^{ {\scriptstyle s } }} $}}
\newcommand{\gpas}
           {\mbox{$G _{_{a}} ^{^{ {\scriptstyle s } }} $}}	           
\newcommand{\gpes}
           {\mbox{$G _{_{E}} ^{^{ {\scriptstyle s } }} $}}
\newcommand{\gpms}
           {\mbox{$G _{_{M}} ^{^{ {\scriptstyle s } }} $}}              
\newcommand{\gatrois}
           {\mbox{$G _{_{A}} ^{^{( 3 )}} $}}
\newcommand{\gahuit}
           {\mbox{$G _{_{A}} ^{^{( 8 )}} $}}
\newcommand{\gad}
           {\mbox{$G _{_{A}} ^{^{D}} $}}              
\newcommand{\gateqzero}
           {\mbox{$ G _{_{ \! a}} ^{^(T=0)} $}}
\newcommand{\gatequn}
           {\mbox{$ G _{_{ \! a}} ^{^(T=1)} $}}
\newcommand{\geteqzero}
           {\mbox{$ G _{_{ \! E}} ^{^(T=0)} $}}
\newcommand{\getequn}
           {\mbox{$ G _{_{ \! E}} ^{^(T=1)} $}}
\newcommand{\gmteqzero}
           {\mbox{$ G _{_{ \! M}} ^{^(T=0)} $}}
\newcommand{\gmtequn}
           {\mbox{$ G _{_{ \! M}} ^{^(T=1)} $}}
\newcommand{\gta}
           {\mbox{$\widetilde G _{_{ \! a}}({Q}^{2}) $}}
\newcommand{\gte}
           {\mbox{$\widetilde G _{_{ \! E}}({Q}^{2}) $}}
\newcommand{\gtm}
           {\mbox{$\widetilde G _{_{ \! M}}({Q}^{2}) $}}
\newcommand{\gtga}
           {\mbox{$\widetilde G _{_{ \! A}}({Q}^{2}) $ }}	
\newcommand{\gtgp}
           {\mbox{$\widetilde G _{_{ \! P}}({Q}^{2}) $ }}            
\newcommand{\gtateqzero}
           {\mbox{$\widetilde G _{_{ \! A}} ^{^(T=0)} $}}
\newcommand{\gtatequn}
           {\mbox{$\widetilde G _{_{ \! A}} ^{^(T=1)} $}}
\newcommand{\gaetequn}
           {\mbox{$G _{_{A}} ^{^{e(T=1)}} $}}

\newcommand{\Jmunc}
           {\mbox{$\widetilde  J_{_{ \! \mu}} ^{^ {NC}} $}}  
\newcommand{\Jmuem}
           {\mbox{$\widetilde  J_{_{ \! \mu}} ^{^ {EM}} $}}      
\newcommand{\Vmus}
           {\mbox{$\widetilde  V_{_{ \! \mu}} ^{^ {(s)}} $}}
\newcommand{\Jmufivenc}
           {\mbox{$\widetilde  J_{_{ \! {\mu 5}}} ^{^ {NC}} $}} 
\newcommand{\Amu}
           {\mbox{$\widetilde  A_{_{ \! \mu}} $}}
\newcommand{\Amueight}
           {\mbox{$\widetilde  A_{_{ \! \mu}} ^{^ {(8)}} $}} 
\newcommand{\Amus}
           {\mbox{$\widetilde  A_{_{ \! \mu}} ^{^ {(s)}} $}}        
       
\newcommand{\rvp}
           {\mbox{$R_{_{V}} ^{^{p}} $}}

\newcommand{\rvn}
           {\mbox{$R_{_{V}} ^{^{n}} $}}
\newcommand{\rap}
           {\mbox{$R_{_{A}} ^{^{p}} $}}
\newcommand{\ratequnsm}
           {\mbox{$R _{_A}^{^{(T=1)SM}}   $}}
\newcommand{\rateqzerosm}
           {\mbox{$R _{_A}^{^{(T=0)SM}}   $}}
\newcommand{\ratequnana}
           {\mbox{$R _{_A}^{^{(T=1)anap}}   $}}
\newcommand{\rateqzeroana}
           {\mbox{$R _{_A}^{^{(T=0)anap}}   $}}

\newcommand{\rvtequn}
           {\mbox{$R _{_V}^{^{T=1}}   $}}
\newcommand{\rvteqzero}
           {\mbox{$R _{_V}^{^{T=0}}   $}}
\newcommand{\razero}
           {\mbox{$R _{_A}^{^{(0)}}   $}}
\newcommand{\rvzero}
           {\mbox{$R _{_V}^{^{(0)}}   $}}
\newcommand{\rateqzero}
           {\mbox{$R _{_A}^{^{T=0}}   $}}
\newcommand{\ratequn}
           {\mbox{$R _{_A}^{^{T=1}}   $}}

\newcommand{\gapqdeux}
           {\mbox{$G _{_A}^{^{p}}({Q}^{2})   $}}

\newcommand{\gvemu}
           {\mbox{$-1 + 4 \sin ^2 \theta _{_W} $}}

\newcommand{\gvuct}
           {\mbox{$\hspace {0.30cm} 1 - \frac{8}{3}\sin ^2 \theta _{_W}$}}

\newcommand{\gvdsb}
           {\mbox{$-1 + \frac{4}{3}\sin ^2 \theta _{_W}$}}

\newcommand{\ustr}{\mbox{$\hspace {0.32cm} \frac{1}{3}$}}
\newcommand{\dstr}{\mbox{$\hspace {0.32cm} \frac{2}{3}$}}
\newcommand{\mustr}{\mbox{$- \frac{1}{3}$}}
\newcommand{\mdstr}{\mbox{$- \frac{2}{3}$}}
\newcommand{\un}{\mbox{$\hspace {0.32cm} 1$}}
\newcommand{\zr}{\mbox{$\hspace {0.32cm} 0$}}
\newcommand{\usd}{\mbox{$\hspace {0.32cm} \frac{1}{2}$}}
\newcommand{\musd}{\mbox{$-\frac{1}{2}$}}
\newcommand{\qtwo}{\mbox{$Q^2$ }}

\def\ssbar{ $<s\bar{s}>$}
\def\scalar{$< N | \bar{s}s | N > $ }
\def\vector{$< N | \bar s\gamma_{\mu}{s} | N > $ }
\def\axial{$< N | \bar s\gamma_{\mu} \gamma_5 {s} | N > $ }
\def\geg{$G_E^\gamma $ }
\def\gmg{$G_E^\gamma $ }
\def\gez{$G_E^Z $ }
\def\gmz{$G_M^Z $ }
\def\gae{$G_A^e $ }
\def\sbar{$\bar{s} $ }
\def\deg{$^{o}$}
\def\gev2{$(GeV/c)^2 $ }
\def\he4{$^4{He}$}

\def\hdiv{\huge \mbox{$ /  \hspace{-0.3cm} $ } \!  \normalsize}
\newcommand{\xu}{\mbox{$x_{_{1}}$}}
\newcommand{\xd}{\mbox{$x_{_{2}}$}}
\newcommand{\sla}[1]
           {\mbox{${#1}\hspace{-0.29cm}/$}}

\newcommand{\mz}{\mbox{$M_{_{Z}}$}}

\maketitle
\section{Introduction}
\label{intro}
Elastic scattering of longitudinally polarized electrons is subject to parity violation through the interference between \gam and  \z0 exchange. These experiments give access to the weak nucleon form factors (FF), which are the equivalent, in the weak sector, of the usual electromagnetic form factors $G_E$ and \gmm. The weak nucleon form factors are related in turn to the strange form factors \ges and \gms, which are the contributions of  strange currents to the form factors (see~\cite{km88} and the following review articles~\cite{ks00,mus94,bh01,bm01}). According to QCD, this strangeness contribution arises from the presence of $s$\sbar pairs in the nucleon sea. 
Many experiments have been performed recently or are still running at Bates (SAMPLE~\cite{bps05,ito04,spa04}), Mainz (PV-A4~\cite{maa04-1,maa04-2}) and Jefferson Lab (\g0~\cite{arm05} and HAPPEX~\cite{ani04}).

Electromagnetic radiation produced from the emission of a real or virtual photon by the electron (incoming or outgoing) or by the target (before or after interaction), gives rises to a radiative tail which extends to very low energies (in theory, down to zero energy for the scattered electron). Since detectors have an experimental resolution and since cuts are used in the data analysis, the measured cross section and asymmetry have to be corrected in order to be compared to theoretical models. 
The first calculations applied to elastic {\sl ep} scattering were done by Tsai~\cite{tsa61}, followed by a series of review papers~\cite{mt69,max69,tsa74}. This formalism has been later extended to scattering of polarized electrons~\cite{mp00}.
All these calculations were done for experiments in which the scattered electron is detected, which was the case of SAMPLE, PV-A4, HAPPEX or \g0 at backward angles~\cite{back}. The originality of the \g0 experiment at forward angles was the detection of recoil protons. In this case, some of the approximations commonly used when scattered electrons are detected, such as the peaking approximation, are no more valid. Thanks to its large mass, radiative emission from the proton is negligible but the proton kinematics is affected by the radiative emission from the electron (angle, energy, \q2).

%\noindent
QED radiative corrections have been calculated for hadronic kinematic variables in $ep$ elastic scattering ~\cite{afa01} and applied to recoil proton polarization. In this case, a method based on an electron structure-function representation, which is the analog of the Drell-Yann representation~\cite{dy70}, was used. These calculations were applied to {\sl ep} scattering experiments done at Jefferson Lab~\cite{jon00}, aiming to determine the ratio of electric to magnetic proton form factors \gep /\gmp  at high momentum transfer as proposed by Akhiezer and Rekalo~\cite{ar74}. The classical method for computing corrections is based on the separation of the momentum phase space into hard- and soft-photons contributions to avoid infrared divergences~\cite{tsa61}. This in\-tro\-duces a cutoff parameter which makes this me\-thod not easily applicable to construct an event ope\-ra\-tor. Bardin and Shumeiko~\cite{bs77} proposed a covariant cancel\-lation me\-thod of infrared divergences which does not introduce additional parameters, however a cutoff has still to be introduced for generating real radiated photons.  

In the present work we calculate the  corrections in both the leptonic and hadronic variables using an original method which is free of infrared divergences. Its interest is  that it is exact and it can be integrated easily into numerical simulation programs such as GEANT. We then apply it to \g0 forward angles using the G0-GEANT code~\cite{g0geant}. Another original feature of the present calculation is the inclusion of \z0 exchange, in addition to  \gam exchange, allowing to calculate the electroweak asymmetry of the radiative tail. We then calculate the corrected quantities of interest: integrated number of counts, time-of-flight (tof) spectra, \q2 distributions. A full account of the present calculations is given in the thesis of Hayko Guler~\cite{gul03} (in French).

In section~\ref{sec:1} we develop the formalism and define the Lagrangians, in section~\ref{sec:4} we describe our method for avoiding divergences, in section~\ref{sec:5} we describe the \g0 apparatus and simulation method. The results are given in section~\ref{sec:6} and we conclude in section~\ref{sec:8}.

\section{Theoretical formalism for elastic scattering}
\label{sec:1}
The data analysis of parity-violating electron-nucleon scattering experiments involve the extraction of an asymmetry
in the helicity-correlated cross section. The raw data are first converted into an
experimentally measured asymmetry $(A_{exp})$. That means that the false asymmetry due to helicity-correlated fluctuations in intensity, energy, positions
and angles of the electron beam have already been taken into account. We assume also that background subtraction has already been done.\\

The asymmetry is commonly defined as:

\begin{eqnarray}\label{eq40a}
A=
\frac{{(\frac{d\sigma}{d\Omega})}^{+}-{(\frac{d\sigma}{d\Omega})}^{-} }
     {{(\frac{d\sigma}{d\Omega})}^{+}+{(\frac{d\sigma}{d\Omega})}^{-} }              
\end{eqnarray} 

\noindent
where ${(\frac{d\sigma}{d\Omega})}^{+}$ et ${(\frac{d\sigma}{d\Omega})}^{-}$ are the cross sections associated with incident electrons having helicity plus and minus respectively. The plus (minus) helicity corresponds to the spin of the electron being aligned and in the same direction as (opposite to) its momentum.  Calculation of the cross section requires the knowledge of the amplitudes which are derived from the currents in the Feynman formalism.

The elastic scattering amplitude has two components corresponding to the electromagnetic part 
$\mathcal{M}_{\bf \gamma}$ and to the weak part $\mathcal{M}_{\bf Z}$

\begin{eqnarray}
 && \hspace*{6mm}
 \mathcal{M} \left ( k' , p', h_{e'}, h_{_{p'}}, 
                     k,   p,  h_{e},  h_{_p}  \right )
\nonumber \\
&&			  
 = \sum_{i=\gamma,Z}^{} \hspace{-2mm}
 \mathcal{M}_{i} \left ( k' , p', h_{e'}, h_{_{p'}}, 
                         k, p, h_{e}, h_{_p}  \right ) 
			\hspace{10mm}
  \label{eq:amp0l_tot}			     
\end{eqnarray}

\noindent
where $ k $ and $ k^{\, \prime} $ are the incident and scattered electron, $ p $ and $ p^{\, \prime} $ are the incident and recoil proton momentum respectively.   $h_{e}, h_{_p} $ and $h_{e'}, h_{_{p'}}$ are the electron and proton helicity in the initial and final state.

\begin{eqnarray}
\label{eq:hadronic-current}
    \mathcal{M}_{\bf \gamma} &=& -ie^{2}~
     ~\frac{1}{q^{2}} ~ 
      {J^{^{P}}}_{ \hspace{-2mm} \nu \hspace{1mm} em}
 \ 
 {j^{\nu} }_{_{\scriptstyle \hspace{-2mm} \, em}}   
  \hspace{7mm}
  \label{eq:M01a}
  \end{eqnarray}
  
 \noindent
  where ${j^{\nu}}_{_{\scriptstyle \hspace{-2mm} \, em} }$ is the Dirac leptonic electromagnetic current:
  
  \begin{eqnarray}
 {j^{\nu}}_{_{\scriptstyle \hspace{-2mm} \, em} }
  &=&
    \bar{u}(k',h_{e'})~\gamma^{\nu}  ~u(k, h_{e})  
  \label{eq:M01b}  
\end{eqnarray}

\noindent
and ${J^{^{P}}}_{ \hspace{-2mm} \nu \hspace{1mm} em}$ is the hadronic part of the electromagnetic  current.
The weak amplitude is given by:

\begin{eqnarray}
 \hspace*{-0mm}
    \mathcal{M}_{\bf Z} \hspace{-0mm} &=& \hspace{-0mm}
    -i   \frac{G}{2\sqrt{2}}~
     ~\frac{1}{1-q^{2}/{\mz^{\! \! 2}}} \hspace{26mm}
     \nonumber \\[1mm]
     &&  \hspace{-1mm}
     \bigg\{  \hspace{3mm}
             \Big(   
                {J^{^P}}_{\hspace{-2mm} \nu \hspace{1mm} nc} + 
                {J^{^P}}_{\hspace{-2mm} \nu \hspace{1mm} nc5}   \Big)
  j^{\nu }_{_{\bf weak}} \
      \nonumber \\[1mm]
     && \hspace{2mm}
             - 	\Big( 
	            {J^{^P}}_{\hspace{-2mm} nc}^{\nu }  +   
	            {J^{^P}}_{\hspace{-2mm} nc5}^{\nu }  \Big) \
		\frac{q_{\nu}q_{\mu}}{\mz^{2}} \
  j^{\mu }_{_{\bf weak}} \ 
     \bigg\}
    \label{eq:M02a}  
 \end{eqnarray}
 
\noindent
and the weak currents are obtained from:               

\begin{eqnarray}
j^{\mu}_{_{ V }} 
&& \hspace{-2mm} =
        g_{V}^{e}  \ \bar{u}(k', h_{e'}) \ \gamma ^\mu \  u(k,h_{e}) 
    \label{eq:M02b}
    \\[1mm]
j^{\mu}_{_{ A }} 
&&\hspace{-2mm}  =
          g_{A}^{e}  \  \bar{u}(k', h_{e'})  \ \gamma ^\mu \gamma ^5 
		    \ u(k,h_{e} )   
    \label{eq:M02c}    	  
  \\[1mm]
j^{\mu}_{_{\bf weak}} && \hspace{-2mm}  = j^{\mu}_{_{ V }} 
 +   j^{\mu}_{_{A }} 
    \label{eq:M02d}     
\end{eqnarray}   

\noindent
where $G$ is the Fermi constant, $g_{V}^{e}$ and $g_{A}^{e}$ are the weak vector and axial charges respectively. For electron scattering and at tree level they reduce to $g_{V}^{e}=\gvemu $ and $g_{A}^{e} = 1$ respectively.

${J^{^P}}_{\hspace{-2mm} \nu \hspace{1mm} nc}$ and ${J^{^P}}_{\hspace{-2mm} \nu \hspace{1mm} nc5}$ are the hadronic weak currents. The hadronic structure is parametrized in terms of form factors:

\vspace{-2mm}
\begin{eqnarray}\label{eq39}
J^{ EM \mu } = < x' | {\hat J}^{ EM \mu } | x > \\
   = \calubar _{x'}  [ \funx (Q^2) \gamma ^\mu +
   i \frac{\fdeuxx (Q^2)}{2 M } \sigma ^{\mu \nu } q_\nu ] \ \calu _{x}
\end{eqnarray}

\vspace{-3mm}
\begin{eqnarray}\label{eq40}
  J^{NC \mu  } =< x' | {\hat J}^{ NC \mu } | x >\\
=
 \calubar _{x'} [ \ftunx  (Q^2)  \gamma ^\mu +
   i \frac{\ftdeuxx (Q^2)}{2 M } \sigma ^{\mu \nu } q_\nu ] \ \calu _{x} 
\end{eqnarray}

\vspace{-3mm}
\begin{eqnarray}\label{eq41}
 J^{ NC \mu 5 } = < x' | {\hat J}^{ NC \mu 5} | x >\\
=
 \calubar _{x'} [ \gtax (Q^2) \gamma ^\mu +
   i \frac{\gtpx (Q^2)}{ M }  q ^\mu ] \gamma ^5 \ \calu _{x}
\end{eqnarray}
 
\noindent
where $ x =p,n$ represents a proton $p$ or a neutron $n$ and
$ \calu _{x} $ and $ \calubar _{x'} $ are the Dirac spinors for the nucleon in the entrance and exit channel respectively. 
$ \funx $ and $ \fdeuxx $ are the electromagnetic form factors,
$ \ftunx $ and $ \ftdeuxx $ are the neutral weak vector form factors and 
$ \gtax $ and $ \gtpx $ are respectively the axial and pseudo-scalar
form factors. The latter enters in the cross section and asymmetry through a 
$\mathcal{M}_{\bf Z}$ squared term which is totally negligible.

\noindent
The observables are usually expressed in terms of the Sachs form factors $\gex$, $\gmx$,  $\gtex$ and  $\gtmx$ rather than the Fermi and Dirac form factors $ \funx$, $\fdeuxx $, $ \ftunx $, $ \ftdeuxx $  :
\begin{eqnarray}
&&
 \hspace{-50mm} \gex (Q^2 ) = \funx (Q^2 ) - \tau \fdeuxx (Q^2 )\\ 
 \gmx (Q^2 ) = \funx (Q^2 )  + \fdeuxx (Q^2 )
\label{eq42}  \\[0.6mm]
&&
 \hspace{-50mm}\gtex (Q^2 ) = \ftunx (Q^2 ) - \tau \ftdeuxx (Q^2 )\\
 \gtmx (Q^2 ) = \ftunx (Q^2 )  + \ftdeuxx (Q^2 )
\label{eq43}  
\end{eqnarray}

\noindent             
where $\tau$ is a kinematic factor defined as $\tau=\frac{{Q}^{2}}{4{{M}_{p}}^{2}}$ and ${M}_{p}$ is the proton mass.

The helicity-correlated cross section is given by:

\begin{eqnarray}
&& 
\frac{d^2 \sigma _{_{\scriptstyle h_e}}}{d\Omega _{e'}}=
 \frac{1}{16 (2\pi )^2 } \; \frac{|\vec k^{\, \prime} |^2  }
                                 {M \,|\vec k |} 
\hspace{40mm}	\nonumber \\[1mm]
&&  \hspace{15mm}				 
 \frac{1}{2} \
 \frac{\sum '|{\mathcal M }|^2 } 
      {\Big|  E_{p'} |\vec k^{\, \prime} |
      + E_{e'} ( | \vec k^{\, \prime} |  - |\vec k | \cos(\theta_{e'}) )
         \Big|}
\label{sec0ehe}	 
\end{eqnarray}

\noindent  
where the summation $\sum '$ is performed over all the 
spin variables except the incident electron helicity. The asymmetry can then be calculated from Eq.~(\ref{eq40a}):

\begin{eqnarray}\label{eq45}
&&
\hspace{-80mm}
{A}_{_{LR}}(\vec{e}N)=-\frac{ {G}_{F}{Q}^{2} }
                            { 4\pi\alpha\sqrt{2}}
             \frac{1}
	          {\varepsilon {(\gex)}^{2} + \tau{(\gmx)}^{2}}\\
 \bigg\{
\varepsilon \, \gex \, \gtex + \tau \, \gmx \, \gtmx
    -(1-4 \, {\sin}^{2}{\theta}_{_W}) \, \varepsilon' \, \gmx \, \gtax
\bigg\}
\end{eqnarray}

 \vspace{2mm}
 \noindent
in which the \qtwo dependence has been omitted for clarity of notation. $\epsilon$, $\epsilon'$ are kinematic factors given by:

\vspace{-3mm}
\begin{eqnarray}\label{eqp7a}
\epsilon=\frac{1}{1+2(1+\tau){\tan}^{2}\frac{\theta_{e'}}{2}}\\
\hspace{1cm}
\epsilon'=\sqrt{\tau(1+\tau)(1-{\epsilon}^{2})}
\end{eqnarray}

\noindent
$\theta_e'$ is the electron scattering angle and ${\theta}_{W}$
is the Weinberg angle.

\noindent
The ultimate purpose of these experiments being to determine the strange content of the 
nucleon, one must isolate the contribution of the $s$-quark in the nucleon form factors. 
In order to do that, we decompose the electromagnetic, neutral and axial currents according to the different flavor contributions $f = u , d, s$:

\begin{eqnarray}
&&
\hspace{-60mm}
< x' |{\bar{u}}_{f}{\gamma}_{\mu}{u}_{f} | x >
\equiv \\
\calubar _{x'} 
\Big( \funfx (Q^2 ) \gamma _\mu  + i \frac{\fdeuxfx (Q^2 )}{2 M }
\sigma _{\mu \nu } q^\nu  \Big) \ \calu _{x}
\label{eq47} \\[0.8mm]
&&
\hspace{-60mm}
< x' |{\bar{u}}_{f}{\gamma}_{\mu}{\gamma}_{5}{u}_{f} | x >
\equiv \\
\calubar _{x'}
\Big(\gtafx  (Q^2 ) \gamma _\mu + i \frac{\ftpfx (Q^2 )  }{ M }  q _\mu
\Big) \, \gamma _5 \, \calu _{x}
\label{eq48}
\end{eqnarray}

\noindent where $u_f$ and $\bar{u}_f$ are the quarks fields.
The pseudo-scalar form factors $  \ftpfx $ being ignored, there are 18 form factors to be evaluated: 9 for the proton and 9 for the neutron. In order to reduce that number we use charge symmetry, assuming that the $p$ and the $n$ are members of a perfect isospin doublet. 
Omitting the  $ Q^2 $-dependence:

\vspace{-0.5cm}

\begin{eqnarray}\label{eq49}
\funu \equiv \funup = \fundn \hspace{1cm}
\fdeuxu \equiv \fdeuxup = \fdeuxdn 
\end{eqnarray}

\vspace{-0.5cm}

\begin{eqnarray}\label{eq50}
\fund   \equiv \fundp = \funun \hspace{1cm}
\fdeuxd \equiv \fdeuxdp = \fdeuxun 
\end{eqnarray}

\vspace{-0.5cm}

\begin{eqnarray}\label{eq51}
\funs   \equiv \funsp = \funsn \hspace{1cm}
\fdeuxs \equiv \fdeuxsp = \fdeuxsn 
\end{eqnarray}

\vspace{-0.5cm}

\begin{eqnarray}\label{eq52}
\gtau \equiv \gtaup = \gtadn\\
\hspace{0.5cm}
\gtad \equiv \gtadp = \gtaun\\
\hspace{0.5cm}
\gtas \equiv \gtasp = \gtasn
\end{eqnarray}

After some algebra~\cite{arv05}, the asymmetry can be finally expressed in terms of the electromagnetic, axial and strange nucleon form factors:

\begin{eqnarray}\label{eq97}
\vspace{-2cm}
&& \hspace*{-5mm}
{A}_{_{LR}}(\vec{e}p)=
-\frac{ {G}_{F}{Q}^{2} }
                            { 4\pi\alpha\sqrt{2}}
\bigg\{ \hspace{2mm}
(1-4{\sin}^{2}{\theta}_{_W})(1+\rvp)
\nonumber \\
&&			  
\hspace{12mm}-(1+\rvn)     
\frac{\varepsilon\gep\gen+\tau\gmp\gmn}
     {\varepsilon{(\gep)}^{2}+\tau{(\gmp)}^{2}}\bigg\}
\nonumber \\[1mm]
&&
\hspace{12mm}+\frac{ {G}_{F}{Q}^{2} }{ 4\pi\alpha\sqrt{2}} 		 
\hspace{2mm}
(1+\rvzero)
\frac{\varepsilon\gep}
             {\varepsilon{(\gep)}^{2}+\tau{(\gmp)}^{2}}\,\ges
\nonumber \\[1mm]
&&
\hspace{12mm}+\frac{ {G}_{F}{Q}^{2} }{ 4\pi\alpha\sqrt{2}} 	     
\hspace{2mm}
(1+\rvzero)
\frac{\tau\gmp}
	     {\varepsilon{(\gep)}^{2}+\tau{(\gmp)}^{2}}\,\gms	     
\nonumber \\[1mm]
&&  \hspace{12mm}
+\frac{ {G}_{F}{Q}^{2} }{ 4\pi\alpha\sqrt{2}} 
      \,\,\,\frac{(1-4{\sin}^{2}{\theta}_{_W}) \, \varepsilon' \gmp}
            {\varepsilon{(\gep)}^{2}+\tau{(\gmp)}^{2}} \, \,\gtap
\end{eqnarray}

\noindent The coefficients \rvp, \rvn and \rvzero are electroweak radiative corrections parameters which can be calculated within the Standard Model ~\cite{pdg04}.

%\noindent
When recoil protons are detected instead of scattered electrons, the helicity-correlated cross section becomes:
 
%the elastic cross section is given by: 
% \begin{eqnarray}
%&&
%\frac{d^2 \sigma}{d\Omega _{p'}}=
% \frac{1}{16 (2\pi )^2 } \; \frac{|\vec p^{\, \prime} |^2  }
%                                 {M \,|\vec k |} 
%\hspace{40mm}	\nonumber \\[1mm]
%&&  \hspace{15mm}
% \frac{1}{4} \
% \frac{\sum |{\mathcal M }|^2 } 
%      {\Big|  E_{e'} |\vec p^{\, \prime} |
%      + E_{p'} ( | \vec p^{\, \prime} |  - |\vec k | \cos(\theta_{p'}) )
%        \Big|}
%\label{sec0p}	
%\end{eqnarray}

\begin{eqnarray}
&&
\frac{d^2 \sigma _{_{\scriptstyle h_e}}}{d\Omega _{p'}}=
 \frac{1}{16 (2\pi )^2 } \; \frac{|\vec p^{\, \prime} |^2  }
                                 {M \,|\vec k |} 
\hspace{40mm}	\nonumber \\[1mm]
&&  \hspace{15mm}				 
 \frac{1}{2} \
 \frac{\sum '|{\mathcal M }|^2 } 
      {\Big|  E_{e'} |\vec p^{\, \prime} |
      + E_{p'} ( | \vec p^{\, \prime} |  - |\vec k | \cos(\theta_{p'}) )
       \Big|}
\label{sec0phe}       
\end{eqnarray}

\noindent The asymmetry calculation follows the same steps as for scattered electron detection.

\section{QED radiative corrections}
\label{sec:4}
\subsection{Parity-violating experiment representation in leptonic variables}
\label{sec:4a}
The aim of the procedure is to get 
a differential cross section $ d^3 \sigma /d\Omega _{e'} dE_{e'} $
without any singularity in the full electron spectrum. We divide the scattered electron energy interval into two regions.
The first one,  
$ E_{e' \, min} \leq E_{e'} \leq E_{e' \, cut}$ where $E_{e' \, cut} \equiv E_{e' \, elas} - \Delta E_{e'}$  corresponds to ``hard photons'' with a minimum energy 
$E_{e' \, min }$ which may be of the order of few MeV. The second one is defined
by $ E_{e' \, cut}  \leq E_{e'}  \leq E_{e' \, elas}$ which
corresponds to the ``soft photon'' region. The maximum energy of the outgoing electron
corresponds to the elastic peak and is denoted by $  E_{e' \, elas} $.
The first requirement is that the integral 
\begin{eqnarray}\label{eqel1}
\int^{E_{e' \, elas}} _{E_{e' \, min}}
&&
 \frac{d^3 \sigma}{d\Omega _{e'} dE_{e'}} \,  dE_{e'}
=
\int^{E_{e' \, cut} } _{E_{e' \, min}}
 \frac{d^3 \sigma}{d\Omega _{e'} dE_{e'}} \,  dE_{e'}
 \nonumber \\
&& 
\hspace{10mm}+
\int^{E_{e' \, elas}} _{ E_{e' \, cut}}
 \frac{d^3 \sigma}{d\Omega _{e'} dE_{e'}} \,  dE_{e'} 
\end{eqnarray}
\noindent
should be as much as possible independent of the cutoff energy $ \Delta E_{e'} $.

\begin{figure}
\resizebox{0.5\textwidth}{!}{
\includegraphics{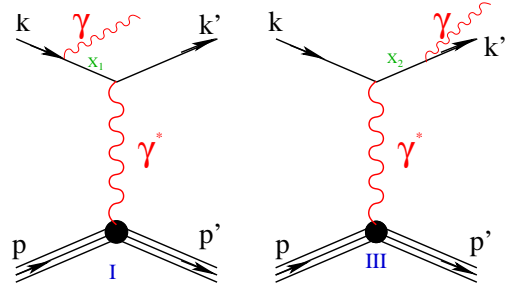}
}
\vspace{-40mm}
\caption{Real-photon emission in virtual-photon exchange}
\label{photon}
\end{figure}

The three-dimensional differential cross section 
 which appears in the first term in the right hand side of
(\ref{eqel1}) is obtained from
the five-dimensional differential cross section: 

\begin{equation}\label{eqel2}
\frac{d^3 \sigma}{d\Omega _{e'} dE_{e'}}
= 
\int \frac{d^5 \sigma}{d\Omega _{e'} dE_{e'} d\Omega _{\gamma}} \ 
d\Omega _{\gamma}
\end{equation}
\noindent
corresponding to the bremsstrahlung process $ e + p \longrightarrow e + p + \gamma $. The two Feynman diagrams describing this process are displayed in
Fig.~\ref{photon}. The integral defined in (\ref{eqel2}) may be calculated 
in the peaking approximation when the scattering angle of the detected electron is
not too small, which is the case for most experiments. In particular, this approximation is very good for the PV-A4 experiment
at forward angle ($30^\circ \leq \theta \leq 40^\circ$) and for the PV-A4
and \g0 experiments at backward angles.
The final result is found to be~\cite{mt69,orv}:

\begin{eqnarray}\label{eqel3}
\frac{d^3 \sigma}{d\Omega _{e'} dE_{e'}}
\approx 
\Big[ \frac{d^3 \sigma}{d\Omega _{e'} dE_{e'}}\Big]_{peaking} \hspace{-2mm}
= && \hspace{0mm}
\mathcal{K}_s \frac{d^2 \sigma _{_0} (E_{e \, s}, Q^2_s )}{d\Omega _{e'} }
\nonumber \\
 &&   \hspace{-3mm}  + 
\mathcal{K}_p \frac{d^2 \sigma _{_0} (E_{e}, Q^2_p)}{d\Omega _{e'} }
\end{eqnarray}
\noindent
where the index $ 0 $ stands for the Born elastic differential cross section.
The term with the $ s $ (respectively $ p $) index represents the contribution of the 
left (right) Feynman diagram of 
Fig.~\ref{photon}. The coefficients 
$ \mathcal{K}_s $ and $\mathcal{K}_p $ are kinematic factors.

\begin{figure}

\resizebox{0.5\textwidth}{!}{
\includegraphics{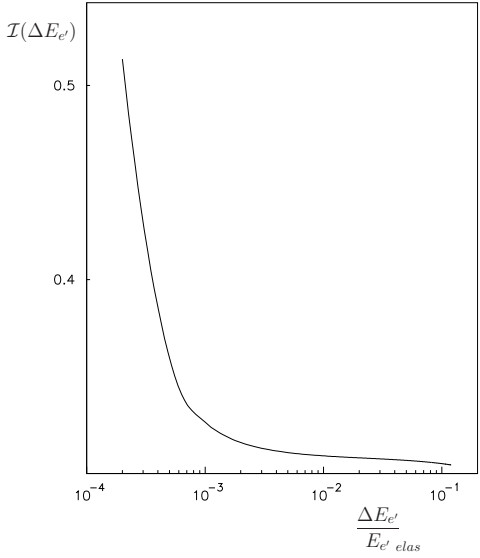}
}
\vspace{-40mm}
    \caption{ Dependence of the right hand side of Eq.~(\ref{eqel1}) with
    $\Delta E_{e'}$ in A4 experiment: $ E_e$=0.855 GeV, 
    $\theta _{e'}$=35$^\circ $}
    \label{cutoffel}
\end{figure}

\vspace{2mm} 
The second term in the right hand side of  (\ref{eqel1})
is usually expressed as 

\begin{eqnarray}\label{eqel4}
%\int^{E_{e' \, elas}} _{ E_{e' \, elas} \hspace{-1mm} - \Delta E_{e'}}
\int^{E_{e' \, elas}} _{ E_{e' \, cut} }
 \frac{d^3 \sigma}{d\Omega _{e'} dE_{e'}} \,  dE_{e'}
 = \! \Big( 1 + \delta (\Delta E_{e'} ) \Big) \ 
 \frac{d^2 \sigma _{_0} (E_{e}, Q^2)}{d\Omega _{e'} }
 \nonumber \\
\end{eqnarray}
\noindent
where the theoretical expression of $ \delta (\Delta E_{e'} )$ is given in ~\cite{mt69,orv}.
The numerical calculation of the right hand side of Eq.~(\ref{eqel1}) is shown
in Fig.~\ref{cutoffel} for the PV-A4 parity-violating experiment.
The minimum value of $ \Delta E_{e'}$ in this kinematics is about 2 MeV.

An analytic expression for the three-dimensional 
differential cross section in the energy range 
$  E_{e' \, cut}  \leq E_{e'} \leq  E_{e' \, elas}$ can be defined as:

\begin{eqnarray}\label{eqel5}
\Big( \frac{d^3 \sigma }{d\Omega _{e'} dE_{e'} }\Bigr) _{anal}
 = a_{_0}(\theta _{e'})\, + \,a_{_1}(\theta _{e'}) \, E_{e'} \, + \, a_{_2}(\theta _{e'}) \, E^2 _{e'}
 \nonumber \\
\end{eqnarray}

\noindent
The three parameters $ a_{_0}$, $ a_{_1}$
and $ a_{_2}$ are fixed using the three conditions: 

\vspace{-2mm}
\begin{eqnarray}
&i)& 
at \ E_{e'} = E_{e' \, cut} :
\nonumber \\
&&   \Big[   \frac{d^3 \sigma}{d\Omega _{e'} dE_{e'}} \Big]_{peaking}
  = \Big( \frac{d^3 \sigma}{d\Omega _{e'} dE_{e'}} \Bigr) _{anal}
  \hspace{10mm}
  \label{eqel6} \\[2mm]
&ii)& 
at \ E_{e'} = E_{e' \, cut}:
\nonumber \\
&& 
 \frac{\partial }{\partial E_{e'} }
   \Big[ \frac{d^3 \sigma}{d\Omega _{e'} dE_{e'}}\Big]_{peaking}  \hspace{-2mm}
   = \hspace{-1mm}
 \frac{\partial }{\partial E_{e'} }   
   \Big( \frac{d^3 \sigma}{d\Omega _{e'} dE_{e'}} \Bigr) _{anal}
  \label{eqel7} \\[2mm] 
& iii) &  
\int ^{E_{e' \, elas}} _{ E_{e' \, cut}} 
\Big(
 \frac{d^3 \sigma}{d\Omega _{e'} dE_{e'}}  \Bigr) _{anal}   \  dE_{e'} 
 = \Big( 1 + \delta (\Delta E_{e'} ) \Big) 
 \nonumber \\
 && \hspace{40mm} \times
 \frac{d^2 \sigma _{_0} (E_{e}, Q^2)}{d\Omega _{e'} }
  \label{eqel8} 
\end{eqnarray}

\begin{figure}

\resizebox{0.45\textwidth}{!}{
\includegraphics{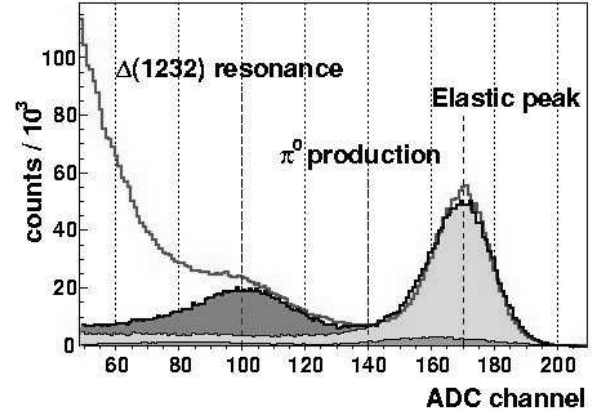}
}
\vspace{-0mm}
    \caption{Comparison of the experimental (red solid line) and simulated spectra (filled grey areas) for the PV-A4 experiment \cite{capozza}. The light-grey area corresponds to the simulated elastic {\it ep} scattering plus radiative corrections.}
    \label{comp_pva4}
\end{figure}

\noindent
Full simulations performed with the Monte-Carlo method and the experimental 
setup in the angular range between $ 40 ^\circ \leq \theta ^\circ \leq 30  $
at $ E_{e} = $ 0.855 GeV \cite{collin,capozza} have shown that the final spectrum is,
within the experimental resolution, independent of the cutoff
parameter $ \Delta E_{e'}$  when its value is  increased by a factor 2 to 4.
The good agreement between the model and the PV-A4 experiment can be seen in Fig.~\ref{comp_pva4}.

The simulated parity-violating asymmetry is then defined as:

\begin{eqnarray}\label{eqel9}
A =
\left\{  \begin{array}{cc}
         A_{elas }    
	&    \hspace{1mm}  E_{e' \, cut}  \leq E_{e'} \leq E_{e' \, elas} 
	\\[2mm]
         \frac{\displaystyle \mathcal{K}_s \sigma _s A_s + \mathcal{K}_p \sigma _p A_p }
	      {\displaystyle \mathcal{K}_s \sigma _s  + \mathcal{K}_p \sigma _p }
	&  \  E_{e' \, min} \leq E_{e'} \leq E_{e' \, cut}    
	 \end{array}	 
\right.
\end{eqnarray}
\noindent
where $\sigma _i \equiv d^2 \sigma _{_0} (E_{e,s}, Q^2_i ) / d\Omega _{e'}$
%$\sigma _i \equiv \frac{d^2 \sigma _{_0} (E_{e}, Q^2_i)}{d\Omega _{e'} }$
, $i=s,p$.
%and $\sigma _p \equiv\frac{d^2 \sigma _{_0} (E_{e}, Q^2_p)}{d\Omega _{e'} }$.
%\sigma _s \equiv d^2 \sigma _{_0} (E_{e \, s}, Q^2_s ) / d\Omega _{e'}$ 
%and 
%$ \sigma _p \equiv d^2 \sigma _{_0} (E_{e}, Q^2_p)/ d\Omega _{e'}$.
 The asymmetries 
$ A_s$ and $ A_p$ are the Born asymmetries calculated
for the kinematics of the $ s $ and $ p $  channels through the 
relations given in the previous section. 

%%%%%%%%%%%%%%%%%%%%%%%%%%%%%%%%%%%%%%%%%%%%%%%%%%%%%%%%%%%%%%%%%%%%%%%%%%%%%%%%%%%%%%

\subsection{Parity-violating experiment in proton variables}
\label{sec:4b}
\begin{figure}
\resizebox{0.5\textwidth}{!}{
\includegraphics{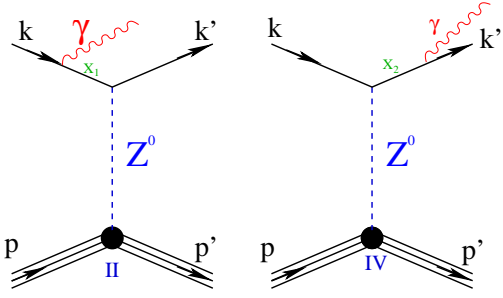}
}
\vspace{-40mm}
    \caption{ Real-photon emission in virtual-\z0 exchange.}
    \label{Z0}
\end{figure}

We describe here the method developed to take into account 
the internal radiative corrections when the proton, instead of the electron, is
detected. Again, we will obtain for the proton spectrum a
differential cross section $ d^3 \sigma /d\Omega _{p'} dE_{p'} $ without
any singularity. The extension of the method derived for the electrons will 
give also the parity-violating asymmetry in the proton channel. 
As in the electron case, we define $ E_{p' \, cut} \equiv E_{p' \, elas} -  \Delta E_{p'}$ 
and we require the integral $ \mathcal{\bf I}(\Delta E_{p'}) $ 
\begin{eqnarray}\label{eqprot0}
&& 
\int^{E_{p' \, elas}} _{E_{p' \, min}}
 \frac{d^3 \sigma}{d\Omega _{p'} dE_{p'}} \,  dE_{p'}
=
\int^{E_{p' \, cut}} _{E_{p' \, min}}
 \frac{d^3 \sigma}{d\Omega _{p'} dE_{p'}} \,  dE_{p'}
 \nonumber \\[1mm]
 && \hspace{10mm}
 +
\int^{E_{p' \, elas}} _{ E_{p' \, cut}}
 \frac{d^3 \sigma}{d\Omega _{p'} dE_{p'}} \,  dE_{p'}
\end{eqnarray}
to be as much as possible independent of the 
energy cutoff $ \Delta E_{p'}$.
 We have to modify the method of the previous section for two reasons.
 First, as we detect the proton, the differential cross section is now given by
  
\begin{equation}\label{eqprot1}
\frac{d^3 \sigma}{d\Omega _{p'} dE_{p'}}
= 
\int \frac{d^5 \sigma}{d\Omega _{p'} dE_{p'} d\Omega _{\gamma}} \ 
d\Omega _{\gamma}
\end{equation}
\noindent
Very forward angles of the outgoing electrons
are allowed when
the integration over all the directions of the photon is performed, so the 
cross section has to be calculated at the amplitude level to be sure that gauge invariance is respected.  
Secondly, as we are interested to correct the experimental asymmetry from the
internal radiative contribution, we need to introduce two more Feynman 
diagrams in the calculation, as shown in Fig.~\ref{Z0}.

The amplitude of the reaction $ e + p \longrightarrow e + p + \gamma $  is 
written as
\vspace{-1mm}
\begin{eqnarray}
 && \hspace*{6mm}
 \mathcal{M} \left ( k' , p', h_{e'}, h_{_{p'}}, p_{\gamma}, 
                        h_{\gamma} ,k, p, h_{e}, h_{_p}  \right )
\nonumber \\
&&			  
 = \sum_{i=\bf I, \cdots IV}^{} \hspace{-2mm}
 \mathcal{M}_{i} \left ( k' , p', h_{e'}, h_{_{p'}}, p_{\gamma}, 
                        h_{\gamma} ,k, p, h_{e}, h_{_p}  \right ) 
			\hspace{10mm}
  \label{eq:ampl_tot}			     
\end{eqnarray}
\noindent
The four-vectors of the exchanged photon and \z0 propagators are
expressed in terms of the kinematic variables:
 
\begin{eqnarray}
 && \hspace*{6mm} q = k - k' - p_\gamma   = p' - p 
\label{eqprot7a}  \\[1mm]
&&  
  x_{_{\scriptstyle 1}} =  k - p_\gamma \hspace{10mm}
  x_{_{\scriptstyle 2}} = k' + p_\gamma = k + p - p'
  \hspace{5mm}
\label{eqprot7b}  
\end{eqnarray}
\noindent
The amplitudes $ \bf I $ and $ \bf III $ due to the exchanged photon 
with the propagator $ - i g_{\nu \nu '}/q^2 $ have one term:

\begin{eqnarray}
    \mathcal{M}_{\bf I} &=& ie^{3}~
     \frac{1}{x^{^{\scriptstyle 2}}_{_{\scriptstyle 1}} -m_e^{2}} 
     ~\frac{1}{q^{2}} ~ 
      {J^{^P}}_{\hspace{-2mm} \nu \hspace{1mm} em}
 \ 
 T^{\nu \mu }_{_{\bf I} \, em}  
  ~{\varepsilon_{\mu}}^{^{\hspace{-1mm} \star}}(p_{\gamma}, h_{\gamma}) 
  \hspace{7mm}
  \label{eq:M1a}  \\[1mm]
 T^{\nu \mu }_{_{\bf I} \, em}
  &=&
    \bar{u}(k',h_{e'})~\gamma^{\nu}  ~  (\sla{\xu} +m_{e})
           \gamma^{\mu}~u(k, h_{e})  
  \label{eq:M1b}  
\end{eqnarray}

\begin{eqnarray}
    \mathcal{M}_{\bf III} &=& ie^{3}~
     \frac{1}{x^{^{\scriptstyle 2}}_{_{\scriptstyle 2}} -m_e^{2}} 
     ~\frac{1}{q^{2}} ~ 
      {J^{^P}}_{\hspace{-2mm} \nu \hspace{1mm} em} 
 \ 
 T^{\mu \nu }_{_{\bf III} \, em}
 ~{\varepsilon_{\mu}}^{^{\hspace{-1mm} \star}}(p_{\gamma}, h_{\gamma}) 
  \hspace{6mm} 
  \label{eq:M3a}  \\[1mm] 
 T^{\mu \nu }_{_{\bf III} \, em} 
 &=&
  \bar{u}(k',h_{e'}) ~ \gamma^{\mu} ~(\sla{\xd} +m_{e})
           \gamma^{\nu}~u(k, h_{e})  
  \label{eq:M3b}  	   
\end{eqnarray}

\noindent
while the amplitudes $ \bf II $ and $ \bf IV $ due to the exchange of the \z0 with the propagator 
$ i ( - g_{\nu \nu '} + q_{\nu} q_{\nu '} /\mz^{2} )/( q^2 - \mz^{2}) $ have each
two different contributions:

\begin{eqnarray}
 \hspace*{-0mm}
    \mathcal{M}_{\bf II} \hspace{-0mm} &=& \hspace{-0mm}
    ie~  \frac{G}{2\sqrt{2}}~
      \frac{1}{x^{^{\scriptstyle 2}}_{_{\scriptstyle 1}} -m_e^{2}} 
     ~\frac{1}{1-q^{2}/{\mz^{\! \! 2}}} \hspace{26mm}
     \nonumber \\[1mm]
     &&  \hspace{-1mm}
     \bigg\{  \hspace{3mm}
             \Big(   
                {J^{^P}}_{\hspace{-2mm} \nu \hspace{1mm} nc} + 
                {J^{^P}}_{\hspace{-2mm} \nu \hspace{1mm} nc5}   \Big)
		\hspace{1mm}
  T^{\nu \mu}_{_{\bf II}} \
 {\varepsilon_{\mu}}^{^{\hspace{-1mm} \star}} (p_{\gamma}, h_{\gamma})
      \nonumber \\[1mm]
     && \hspace{2mm}
             - 	\Big( 
	            {J^{^P}}_{\hspace{-2mm} nc}^{\nu '}  +   
	            {J^{^P}}_{\hspace{-2mm} nc5}^{\nu '}  \Big) \
		\frac{q_{\nu}q_{\nu'}}{\mz^{2}} \
  T^{\nu \mu}_{_{\bf II}} \
  {\varepsilon_{\mu}}^{^{\hspace{-1mm} \star}} (p_{\gamma}, h_{\gamma}) 
     \bigg\}
    \label{eq:M2a}  
  \\[1mm]
T^{\nu \mu}_{_{{{\bf II}} \, V }} 
&& \hspace{-4mm} =
        g_{V}^{e}  \ \bar{u}(k', h_{e'}) \ \gamma ^\nu 
		    (\sla{\xu} +m_{e}) \gamma ^\mu \  u(k,h_{e} )
    \label{eq:M2b}
    \\[1mm]
T^{\nu \mu}_{_{{\bf II} \, A }} 
&&\hspace{-4mm}  =
          g_{A}^{e}  \  \bar{u}(k', h_{e'})  \ \gamma ^\nu \gamma ^5 
		    (\sla{\xu} +m_{e}) \gamma ^\mu  \ u(k,h_{e} )   
    \label{eq:M2c}    	  
  \\[1mm]
T^{\nu \mu}_{_{\bf II}} && \hspace{-4mm}  = T^{\nu \mu}_{_{{\bf IV} \, V }} 
 +   T^{\nu \mu}_{_{{\bf IV} \, A }}, 
    \label{eq:M2d}     
\end{eqnarray}   

\begin{eqnarray}
 \hspace*{-4mm}
    \mathcal{M}_{\bf IV} \hspace{-0mm} &=& \hspace{-0mm}
    ie~  \frac{G}{2\sqrt{2}}~
      \frac{1}{x^{^{\scriptstyle 2}}_{_{\scriptstyle 2}} -m_e^{2}} 
     ~\frac{1}{1-q^{2}/{\mz^{\! \! 2}}} \hspace{26mm}
     \nonumber \\[1mm]
     &&  \hspace{-1mm}
     \bigg\{  \hspace{3mm}
             \Big(   
                {J^{^P}}_{\hspace{-2mm} \nu \hspace{1mm} nc} + 
                {J^{^P}}_{\hspace{-2mm} \nu \hspace{1mm} nc5}   \Big)
		\hspace{1mm}
 T^{\mu \nu}_{_{\bf IV}}
 \ {\varepsilon_{\mu}}^{^{\hspace{-1mm} \star}} (p_{\gamma}, h_{\gamma})
      \nonumber \\[1mm]
     && \hspace{2mm}
             - 	\Big( 
	            {J^{^P}}_{\hspace{-2mm} nc}^{\nu '}  +   
	            {J^{^P}}_{\hspace{-2mm} nc5}^{\nu '}  \Big) \
		\frac{q_{\nu}q_{\nu'}}{\mz^{2}}  \
  T^{\mu \nu}_{_{\bf IV}} \
 {\varepsilon_{\mu}}^{^{\hspace{-1mm} *}} (p_{\gamma}, h_{\gamma}) 
     \bigg\}
    \label{eq:M4a}  
  \\[1mm]
T^{\mu \nu}_{_{{\bf IV} \, V }} 
&& \hspace{-4mm}  =
        g_{V}^{e}  \ \bar{u}(k', h_{e'}) \ \gamma ^\mu 
		    (\sla{\xd} +m_{e}) \gamma ^\nu \  u(k,h_{e} )
    \label{eq:M4b}
    \\[1mm]
T^{\mu \nu}_{_{{\bf IV} \, A }} 
&&\hspace{-4mm}   =
          g_{A}^{e}  \  \bar{u}(k', h_{e'})  \ \gamma ^\mu  
		    (\sla{\xd} +m_{e}) \gamma ^\nu  \gamma ^5  \ u(k,h_{e} )   
    \label{eq:M4c}
   \\[1mm]
T^{\mu \nu}_{_{\bf IV}} && 
 \hspace{-4mm}   = T^{\mu \nu}_{_{{\bf VI} \, V }} +
       	      T^{\mu \nu}_{_{{\bf VI }\, A }}
    \label{eq:M4d}	      
\end{eqnarray}

In the energy range where the parity-violating experiments are performed 
($ 0.1 \leq Q^2 \leq 1  $\gev2,
 terms proportional to $ 1/{\mz^{2}} $ are neglected, therefore:

\begin{eqnarray}
 \hspace*{-0mm}
    \mathcal{M}_{\bf II} \hspace{-0mm} &\approx& \hspace{-0mm}
    ie~  \frac{G}{2\sqrt{2}}~
      \frac{1}{x^{^{\scriptstyle 2}}_{_{\scriptstyle 1}} -m_e^{2}} 
     \hspace{25mm}
     \nonumber \\[1mm]
     &&  
     \bigg\{  \hspace{0mm}
             \Big(   
                {J^{^P}}_{\hspace{-2mm} \nu \hspace{1mm} nc} + 
                {J^{^P}}_{\hspace{-2mm} \nu \hspace{1mm} nc5}   \Big)
		\hspace{0mm}
 \Big(   T^{\nu \mu}_{_{{\bf II} \, V }} + T^{\nu \mu}_{_{{\bf IV} \, A }} \Big) \		   	      
{\varepsilon_{\mu}}^{^{\hspace{-1mm} \star}} (p_{\gamma}, h_{\gamma}) 
     \bigg\}
     \nonumber \\	  
     \label{eq:M2ap}	   
\end{eqnarray}
\noindent
and
\begin{eqnarray}
 \hspace*{-0mm}
    \mathcal{M}_{\bf IV} \hspace{-0mm} &\approx & \hspace{-0mm}
    ie~  \frac{G}{2\sqrt{2}}~
      \frac{1}{x^{^{\scriptstyle 2}}_{_{\scriptstyle 2}} -m_e^{2}} 
     \hspace{25mm}
     \nonumber \\[1mm]
     &&  
     \bigg\{  \hspace{0mm}
             \Big(   
                {J^{^P}}_{\hspace{-2mm} \nu \hspace{1mm} nc} + 
                {J^{^P}}_{\hspace{-2mm} \nu \hspace{1mm} nc5}   \Big)
		\hspace{0mm}
 \Big(   T^{\mu \nu}_{_{{\bf IV }\, V }} + T^{\mu \nu}_{_{{\bf VI} \, A }} \Big) \			        
	      {\varepsilon_{\mu}}^{^{\hspace{-1mm} \star}} (p_{\gamma}, h_{\gamma}). 
     \bigg\}
     \nonumber \\	  
     \label{eq:M4ap}	   
\end{eqnarray}

The total amplitude is the sum of two terms 

\begin{eqnarray}\label{eqp7a_2}
    \mathcal{M} = \mathcal{M}^{^{\scriptstyle PC}} + \mathcal{M}^{^{\scriptstyle PV}}
\end{eqnarray}
\noindent
The interference of these two terms will produce the parity-violating asymmetry. The Parity Conserving 
amplitude $ \mathcal{M}^{^{\scriptstyle PC}}$ is due to photon exchange and it contains a part
of the \z0 exchange. The Parity Violating  amplitude $ \mathcal{M}^{^{\scriptstyle PV}}$ 
is due to part of the \z0 exchange contribution in the Feynman diagrams $\bf II$ and $\bf IV$. 
Explicitly, this amplitudes is 
 
\begin{eqnarray}\label{eqprot8}
    \mathcal{M}^{^{\scriptstyle PV}} = \mathcal{M}^{^{\scriptstyle PV}}_{\bf II} 
                        + \mathcal{M}^{^{\scriptstyle PV}}_{\bf IV}
\end{eqnarray}

\begin{eqnarray}
\hspace*{-0mm}
    \mathcal{M}^{^{\scriptstyle PV}}_{\bf II} \hspace{-0mm} &=& \hspace{-0mm}
    ie~  \frac{G}{2\sqrt{2}}~
      \frac{1}{x^{^{\scriptstyle 2}}_{_{\scriptstyle 1}} -m_e^{2}} 
      \hspace{45mm}
      \nonumber \\[1mm]
      && \hspace{0mm}
     \Big(  \hspace{0mm}        
     {J^{^P}}_{\hspace{-2mm} \nu \hspace{1mm} nc} \hspace{1.5mm}
     T^{\nu \mu}_{_{{\bf II} \, A }}	\              
        +  \ {J^{^P}}_{\hspace{-2mm} \nu \hspace{1mm} nc5} \
     T^{\nu \mu}_{_{{\bf II} \, V }}    
    \hspace{0mm} \Big) \
    	{\varepsilon_{\mu}}^{^{\hspace{-1mm} \star}} (p_{\gamma}, h_{\gamma})   
     \label{eq:M2pva}		
\end{eqnarray}

\begin{eqnarray}
\hspace*{-0mm}
    \mathcal{M}^{^{\scriptstyle PV}}_{\bf IV} \hspace{-0mm} &=& \hspace{-0mm}
    ie~  \frac{G}{2\sqrt{2}}~
      \frac{1}{x^{^{\scriptstyle 2}}_{_{\scriptstyle 2}} -m_e^{2}} 
      \hspace{45mm}
      \nonumber \\[1mm]  
      && \hspace{0mm}          
     \Big(  \hspace{0mm}        
     {J^{^P}}_{\hspace{-2mm} \nu \hspace{1mm} nc} \hspace{1.5mm}
     T^{\mu \nu}_{_{{\bf IV} \, A }}	\              
        +  \ {J^{^P}}_{\hspace{-2mm} \nu \hspace{1mm} nc5} \
     T^{\mu \nu}_{_{{\bf IV} \, V }}    
    \hspace{0mm} \Big) \
    	{\varepsilon_{\mu}}^{^{\hspace{-1mm} *}} (p_{\gamma}, h_{\gamma})	  
     \label{eq:M4pva}		
\end{eqnarray}
The differential cross section is then calculated in the laboratory system in terms 
of the amplitudes by:

\begin{eqnarray}\label{eqprot9}
&&
\frac{d^5 \sigma}{d\Omega _{p'} dE_{p'} d\Omega _{\gamma}} =
 \frac{1}{32 (2\pi )^5 } \; \frac{|\vec p^{\, \prime} | E_\gamma }
                                 {M \,|\vec k |} 
\hspace{30mm} \nonumber \\[1mm]
&& \hspace{25mm} 				 
 \frac{1}{4} \
 \frac{\sum |{\mathcal M }|^2 } 
      { | E_\gamma + E_{e'} + 
        \vec{u}_\gamma \cdot ( \vec p^{\, \prime}  - \vec k ) |}
\end{eqnarray}
\noindent
where the summation is performed over all the helicity states of the
incoming electron, the target, the outgoing proton and the outgoing photon. The 
differential cross section of the outgoing proton is then expressed as in (\ref{eqprot1}), 
after integration
over all the photon angles. 

The parity-violating asymmetry is calculated in a similar way.
First we calculate the differential cross section as a function of the
beam helicity $ h_e = \pm 1/2 $ :

\begin{eqnarray}\label{eqprot10}
&&
\frac{d^5 \sigma _{_{\scriptstyle h_e}}}
     {d\Omega _{p'} dE_{p'} d\Omega _{\gamma}} =
 \frac{1}{32 (2\pi )^5 } \; \frac{|\vec p^{\, \prime} | E_\gamma }
                                 {M \,|\vec k |} 
\hspace{30mm} \nonumber \\[1mm]
&& \hspace{25mm}				 
 \frac{1}{2} \
 \frac{\sum '|{\mathcal M }|^2 } 
      { | E_\gamma + E_{e'} + 
        \vec{u}_\gamma \cdot ( \vec p^{\, \prime}  - \vec k ) |}
\end{eqnarray}
\noindent
The prime index over the summation means that the sum is performed over all the 
spin variables except the incident electron helicity. The parity-violating asymmetry
of the proton spectrum then reads:

\begin{eqnarray}\label{eqprot11}
&&
A \! = \! \bigg(     
            \frac{d^3 \sigma _{_{1/2}}}{d\Omega _{p'} dE_{p'}}
           -\frac{d^3 \sigma _{_{-1/2}}}{d\Omega _{p'} dE_{p'}}
    \bigg) \!  
    \hdiv  
    \bigg(    	   	   
	 {  \frac{d^3 \sigma _{_{1/2}}}{d\Omega _{p'} dE_{p'}}
	   +\frac{d^3 \sigma _{_{-1/2}}}{d\Omega _{p'} dE_{p'}} }
    \bigg)	  
    \nonumber \\    
\end{eqnarray}
\noindent
with
\begin{eqnarray}\label{eqprot12}
\frac{d^3 \sigma  _{_{\scriptstyle h_e}} }{d\Omega _{p'} dE_{p'}}
= 
\int \frac{d^5 \sigma  _{_{\scriptstyle h_e}}}
          {d\Omega _{p'} dE_{p'} d\Omega _{\gamma}} \ 
d\Omega _{\gamma}
\end{eqnarray}
Now we are able to calculate the integral 
\begin{eqnarray}\label{eqprot13}
\int^{E_{p' \, cut}} _{E_{p' \, min}}
 \frac{d^3 \sigma}{d\Omega _{p'} dE_{p'}} \,  dE_{p'}
\end{eqnarray}
\noindent
for any value of $ \Delta E_{p'} \not= 0 $. As in the electron case, the integral
in the energy range 
$  E_{p' \, cut}  \leq E_{p'}  \leq E_{p' \, elas}$ is       
proportional
to the Born elastic differential cross section:

\begin{eqnarray}\label{eqprot14}
\int^{E_{p' \, elas}} _{ E_{p' \, cut}}
 \frac{d^3 \sigma}{d\Omega _{p'} dE_{p'}} \,  dE_{p'}
 = \mathcal{A}(\Delta E_{p'}) \ 
 \frac{d^2 \sigma _{_0} (E_{e}, Q^2)}{d\Omega _{p'} } 
 \hspace{5mm}
\end{eqnarray}
\noindent
Its calculation is given by the following ratio
\begin{eqnarray}\label{eqprot15}
&&
\mathcal{A}(\Delta E_{p'}) =
\hspace{2mm}
\bigg(
\int  \mathcal{K}(\Delta E_{p'}) \ 
\frac{d^5 \sigma}{d\Omega _{p'} dE_{p'} d\Omega _{\gamma} } 
\ d\Omega _{\gamma}
\bigg)
\hspace{10mm} \nonumber \\[1mm]
&& \hspace{18mm}
\hdiv
\bigg(
\int  
\frac{d^5 \sigma}{d\Omega _{p'} dE_{p'} d\Omega _{\gamma}} 
\ d\Omega _{\gamma}
\bigg)
\end{eqnarray}
\noindent
with~\cite{van00}
\begin{eqnarray}\label{eqprot16}
\mathcal{K}(\Delta E_{p'}) \equiv
\frac{e^{\delta _{vertex} + \delta _{R} }}
          {( 1 - \delta _{vacuum}/2 )^2 }
\end{eqnarray}
\noindent
The meaning of $\mathcal{K}(\Delta E_{p'})  $ is clear. For each value of
 $ \theta _{p'} $, $ \phi _{p'} $, $  E_{p'}$, $\theta _{\gamma} $ and
 $ \phi _{\gamma} $, the value of $ \Delta E_{p'}$ is equal to
 $ E_{p' \, elas} - E_{p'}$. The three body kinematics give
 the energy of the photon $ E_\gamma $ and the complete kinematics of the 
 outgoing
 electron  $ \theta _{e'}$, $ \phi _{e'}$ and $ E_{e'}$ 
  through the energy-momentum conservation. 
 Comparison with the elastic scattering
 $ e + p \rightarrow e+p $ reaction at the same angle gives the value of 
 $ \Delta E_{e'} = E_{e' \, elas} -E_{e'} $. The ratio 
 $ \mathcal{K}=e^{\delta _{vertex} + \delta _{R} }/
          ( 1 - \delta _{vacuum}/2 )^2  $ is the attenuation factor,
 which depends on $ \Delta E_{e'}$,	   
 induced by the internal radiative correction on the  electron side. It is 
 a generalization to all orders of the $ 1 + \delta $ term 
 of Eq.~(\ref{eqel8})
 as it can be seen if we make a Taylor
 expansion of $ \mathcal{K} $. Finally the attenuation
 factor $ \mathcal{A}(\Delta E_{p'}) $ as defined in equation
 (\ref{eqprot15}) is the average attenuation factor when we integrate
 over all the directions of the photon. The explicit formulae used in the code 
 to calculate $ \delta _{vertex}$, $ \delta _{R}$  and 
 $ \delta _{vacuum}$ are taken from~\cite{van00} :
 
\begin{eqnarray} 
\delta _{R}  =  
\frac{\alpha}{\pi} 
\bigg\{ \hspace{-4mm} &&
ln \Big( \frac{ (\Delta E_{s} )^2}{E_{e} E_{e'}} \Big)
\Big[ \, ln \Big( \frac{Q^2}{m^2 _e} \Big) -1 \Big]
- \frac{1}{2} ln^2 \Big( \frac{E_{e}}{E_{e'}}   \Big)
\nonumber \\[1mm]
&&\hspace{-3mm}
+ \frac{1}{2} ln^2 \Big( \frac{Q^2}{m^2 _e}   \Big)
-\frac{\pi ^2}{3} + Sp \Big(\cos ^2 (  \theta _{e'}/2 ) \Big) \
\bigg\}  \hspace{0mm}
 \label{eqprot17}          \\[2mm] 
\Delta E_{s} & \hspace{-0mm} =  & \hspace{-0mm}
\frac{E_e}{E_{e' \, elas}  } ( E_{e' \, elas} - E_{e'} )
\label{eqprot18}           \\[1mm]
\delta _{vacuum} & \hspace{-0mm} =  & \hspace{-0mm}
\frac{\alpha}{\pi} \frac{2}{3} 
 \bigg\{ - \frac{5}{3} + ln \Big( \frac{Q^2}{m^2 _e}   \Big) \bigg\}
\label{eqprot19}           \\[1mm] 
\delta _{vertex} & \hspace{-0mm} =  & \hspace{-0mm}
\frac{\alpha}{\pi}
 \bigg\{
\frac{3}{2}  ln \Big( \frac{Q^2}{m^2 _e}   \Big) -2 
- \frac{1}{2} ln^2 \Big( \frac{Q^2}{m^2 _e}   \Big)
 + \frac{\pi ^2}{6}
 \bigg\} 
\label{eqprot20} 
\end{eqnarray}

\begin{figure}
\resizebox{0.45\textwidth}{!}{ 
\includegraphics{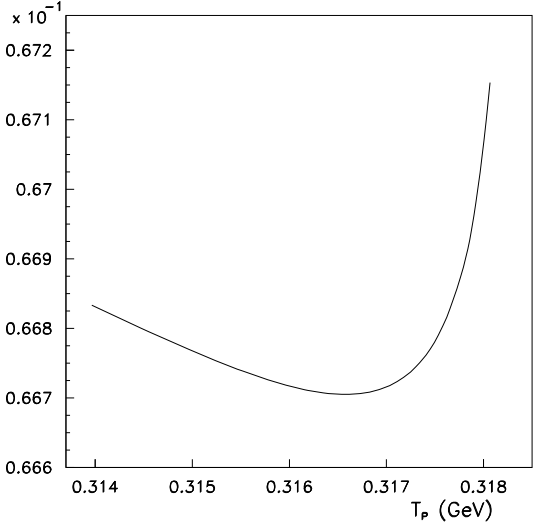}
}
    \caption{ $ \mathcal{I}(\Delta E_{p'} )$ as a function of the kinetic 
    energy of the detected proton  for 
    $\theta _{p'}$=60$^\circ $}
    \label{cutoffprot}
\end{figure}

 The value of the integral 
  $\mathcal{\bf I}(\Delta E_{p'} )  $ as a function of $ \Delta E_{p'}$ 
  has been performed  for
   $  48^\circ  \leq \theta _{p'} \leq 77^\circ $. It is  plotted
  in Fig.~\ref{cutoffprot} for one scattering angle of the detected proton.
  The value of the cutoff parameter is chosen so that this integral reaches
  its minimum value. 
 
As in the electron case, we assume that for the kinetic energy range of the scattered proton
$   T_{p' \, elas} - \Delta E_{p'}  \leq T_{p'}  \leq T_{p' \, elas}$ 

\begin{eqnarray}\label{eqprot21}
&&
 \Big( \frac{d^3 \sigma }{d\Omega _{p'} dE_{p'} }\Bigr) _{anal}
 = a_{_0}(\theta _{p'})\, + \,a_{_1}(\theta _{p'}) 
 \, (T_{p'} - T_{p' \, elas} ) 
 \nonumber \\
 && \hspace{30mm}
 + \, a_{_2}(\theta _{p'}) \, (T_{p'} - T_{p' \, elas} ) ^2
\end{eqnarray}
\noindent
The determination of the three coefficients $ a_{_0}$, $ a_{_1}$ and $a_{_2} $ is 
obtained by the following conditions:

\begin{eqnarray}
&i)& {\mbox{at}} \ E_{p'} = E_{p' \, cut} :
\nonumber \\
&&   \Big(   \frac{d^3 \sigma}{d\Omega _{p'} dE_{p'}} \Big)
  = \Big( \frac{d^3 \sigma}{d\Omega _{p'} dE_{p'}} \Bigr) _{anal}
  \hspace{10mm}
  \label{eqprot22} \\[2mm]
&ii)& 
{\mbox{at}} \ E_{p'} = E_{p' \, cut}:
\nonumber \\
&& 
 \frac{\partial }{\partial E_{p'} }
   \Big( \frac{d^3 \sigma}{d\Omega _{p'} dE_{p'}}\Big)  \hspace{-0mm}
   = \hspace{-0mm}
 \frac{\partial }{\partial E_{p'} }   
   \Big( \frac{d^3 \sigma}{d\Omega _{p'} dE_{p'}} \Bigr) _{anal}
  \label{eqprot23} \\[4mm]
& iii) &  
\int ^{E_{p' \, elas}} _{ E_{p' \, cut}} 
\Big(
 \frac{d^3 \sigma}{d\Omega _{p'} dE_{p'}}  \Bigr) _{anal}   \  dE_{p'} 
 =  
 \mathcal{A}(\Delta E_{p'}) \  
 \nonumber \\
 && \hspace{20mm} \times
 \frac{d^2 \sigma _{_0} (E_{e}, Q^2)}{d\Omega _{p'} }
  \label{eqprot24} 
\end{eqnarray}

\begin{figure}
\resizebox{0.45\textwidth}{!}{ 
\includegraphics{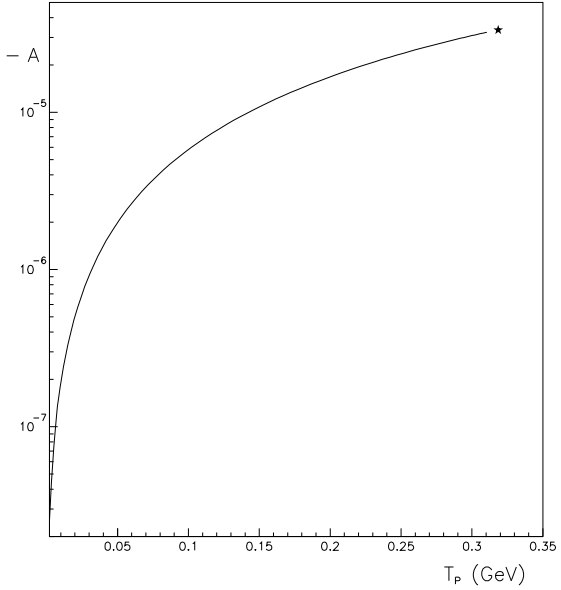}
}
\vspace{-1mm}
    \caption{ Parity-violating asymmetry as a function of the kinetic energy of the detected proton  for $\theta _{p'}$=60$^\circ $. The star $\star$ represents the Born asymmetry.}
    \label{asymprot}
\end{figure}

The parity-violating asymmetry is calculated through the relation
(\ref{eqprot11})  for $ E_{p'} \leq E_{p' \, cut} $ and near the elastic peak, 
its value is 
linearly interpolated between its value at  
$ E_{p'} =E_{p' \, cut}$ and the Born asymmetry 
calculated at
 $ E_{p'} =E_{p' \, elas}$. The variation of this asymmetry as a function of
  the kinetic
energy of the scattered proton is plotted in Fig.~\ref{asymprot} for
one angle.

\section{Application to the \g0 experiment at forward angles}
\label{sec:5}
The \g0 experiment~\cite{G0equip}, performed in Hall C at Jefferson Lab, measures the parity-violating elastic electron scattering from the nucleon.  
Asymmetries of the order of one part per million from scattering of a polarized electron beam are determined 
using a dedicated apparatus.  It consists of specialized beam monitoring and control systems, a 
cryogenic hydrogen target and a superconducting, toroidal magnetic spectrometer equipped with plastic 
scintillation counters as well as fast readout electronics for the measurement 
of individual events. 

In the forward-angle configuration, a polarized electron beam of 40 $\mu A$ with an energy of $3.031 \pm 0.001$ GeV was used over the measurement period of 700 h. 
It was generated by a strained GaAs polarized source~\cite{poelker00a} with 32-ns pulse timing (rather than the standard 2 ns) to allow for time-of-flight (tof) measurements.  The average beam polarization, measured with a M\o ller polarimeter~\cite{hauger99} in interleaved runs, was $73.7 \pm 1.0\%$.
Helicity-correlated current and position changes were corrected with active feedback to levels of about 0.3 parts-per-million (ppm) and 10 nm, respectively.  Corrections to the measured asymmetry were applied via linear regression for residual helicity-correlated beam current, position, angle and energy variations, and amounted to a negligible total of 0.02 ppm; the largest correction was 0.01 ppm for helicity-correlated current variation.

\begin{figure}
\resizebox{19.pc}{!}{\includegraphics{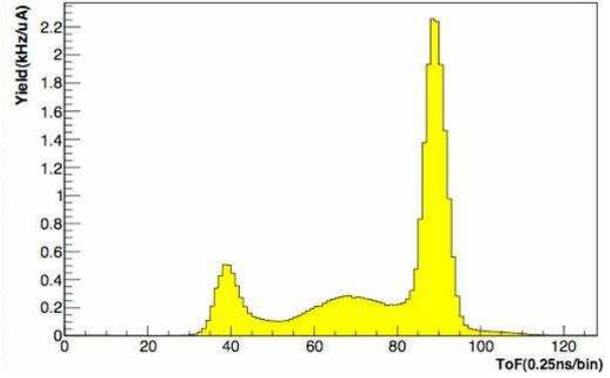}}
\caption{\label{fig:tof} Experimental yield as a function of tof for detector 8. The elastic protons correspond to the rightmost peak.}
\end{figure}

The polarized electrons scattered from a 20 cm liquid hydrogen target; the recoiling elastic protons were detected to allow simultaneous measurement of the wide range of mo\-men\-tum trans\-fer $0.12 \le Q^2 \le 1.0$\gev2.\newline  This was achieved using a novel toroidal spectrometer designed to measure the entire range with a single field setting and with precision comparable to previous experiments. The spectrometer included an eight-coil superconducting magnet and eight sets (or octants) of scintillating detectors.  
Four octants (numbered 1-3-5-7) and their associated electronics were built by the North-American (USA, Canada) part of the \g0 collaboration and four octants (2-4-6-8) and their associated electronics were built by the French (IPN Orsay, LPSC Grenoble) part of the \g0 collaboration. 
Each set consisted of 16 scintillator pairs used in coincidence to cover the range of momentum transfers (smallest detector number corresponding to the lowest momentum transfer). The scattering angle varies from 52 to 76 degrees, depending on detector number.  Because of the correlation between the momentum and scattering angle of the elastic protons (higher momentum corresponds to more forward proton scattering angles), detector 15 covered the range of momentum transfers between 0.44 and 0.88 \gev2, which we divided into three tof bins with average momentum transfers of 0.51, 0.63 and 0.79 \gev2.  For the same reason, detector 14 had two elastic peaks separated in tof with momentum transfers of 0.41 and 1.0 \gev2; detector 16, used to determine backgrounds, had no elastic acceptance.  Custom time-encoding electronics
sorted detector events by tof; elastic protons arrived about 20 ns after the passage of the electron bunch through the target. A typical time-of-flight spectrum is shown in  Fig.~\ref{fig:tof}.  The spectrometer field integral and ultimately the $Q^2$ calibration ($\Delta Q^2/Q^2 = 1$\%) was fine-tuned using the measured tof difference between pions and elastic protons for each detector.
All rates were corrected for dead-times of $10-15\%$ on the basis of the measured yield dependence on beam current; the corresponding uncertainty in the asymmetry is $\sim 0.05$ ppm.  The final results of the \g0 forward-angle experiment are shown in~\cite{arm05}.

Radiative corrections for \g0 have been estimated in a simulation using the G0-GEANT package~\cite{g0geant}. 
The electron can, in principle, loose all its energy through
radiation, but the probability that it looses 500 MeV or less is 96\%. 
Moreover, 60\% loose 1 keV or less. The few events for which the electron energy loss is more than 500 MeV correspond to proton having times of flight out of the \g0 experimental cuts, thus they are not considered in our calculation.
Therefore the GEANT simulations
have been done in the energy interval E$_{inc}$= 2.5-3.0 GeV only and for recoil proton angle 
$\theta _{p'}=48^o-77^o$ and energy  T$_{p'}$=2
MeV-T$_{p'}^{el}$ . The cross sections have been interpolated for
intermediate values using a spline method.
In order to obtain rates, each event (number $j$) is normalized through a weight w$_j$ proportional to the cross section\cite{vm05}:

\begin{eqnarray}
  \label{eq:norm_gal}
  w_j=\mathcal{L}\frac{\Delta\phi}{N_{T}}\frac{d^3\sigma_{j}}{d\Omega dE}\sin{\theta_{j}}
  [\theta_{max}(E^{inc}_j)-\theta_{min}(E^{inc}_j)] 
  \nonumber \\[1mm]
&& \hspace{-60mm}
  [E_{max}(E^{inc}_j,\theta_j)-E_{min}(E^{inc}_j,\theta_j)]
\end{eqnarray}

\noindent
where $\mathcal{L}$ is the luminosity, $\Delta \Phi $ is the polar angle opening
and N$_T$ is the number of drawings. In the case of elastic scattering (Born term), the weight is simply given
by

\hspace{2mm}
$ w_j=\mathcal{L}\frac{\Delta\theta\Delta\phi}{N_{T}}\sin{\theta_{j}}\frac{d^2\sigma_{j}}{d\Omega}$
  
\section {Results}
\label{sec:6}
\subsection {Time-of-flight spectra}
\label{sec:6a}
\begin{figure}
\resizebox{0.49\textwidth}{!}{ 
\includegraphics{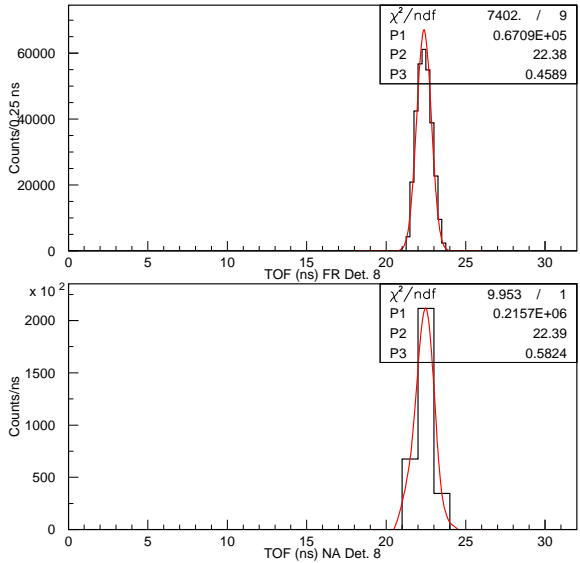}
}
\vspace{-1mm}
    \caption{ Simulated elastic-proton time-of-flight distributions for Det. 8 (top: FR, bottom: NA). The gaussian fits are performed to extract the position of the mean.}
    \label{fits_french}
\end{figure}

Two calculations are performed without and with RC:

\hspace{1cm}- in the first case, the time-of-flight of elastically scattered protons, without any energy loss nor radiative corrections is calculated. The width of the peak is essentially given by the experimental resolution as calculated by GEANT. The elastic peak is fitted with a Gaussian allowing to determine the position of the maximum, in order to define cuts within which the asymmetry is calculated. The resulting spectra are shown in Fig.~\ref{fits_french}, where only detector 8, corresponding to the middle of the focal plane, is shown for reference. 
The only difference in the two spectra is a binning of 250 ps for the  French (FR) electronics (top) and a binning of 1 ns for the North-American (NA) electronics (bottom). 
%%%%%%%%%%%%%%%%%%%%%%%%%%%%%%%%%%%%%%%%%%%%%%%%%%%%%

\begin{figure}
\resizebox{0.5\textwidth}{!}{ 
\includegraphics{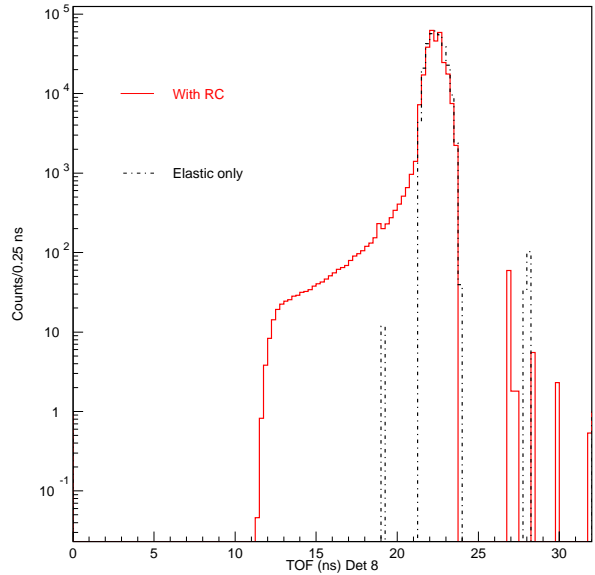}
}
\vspace{-1mm}
    \caption{ Comparison of time-of-flight spectra (FR Det. 8) without (black dash-dotted line) and with (red/grey solid line) radiative corrections.}
    \label{tof_comparison}
\end{figure}

\hspace{1cm}- in the second case, the proton tof spectra are calculated after applying energy losses and full RC. The result obtained for detector 8 is shown in red (grey) in Fig.~\ref{tof_comparison}, overlaid to the pure elastic spectra (in black). 
%%%%%%%%%%%%%%%%%%%%%%%%%%%%%%%%%%%%%%%%%%%%%%%%%%%%%%%%%%%%%

\noindent
One should notice the following
paradox: Inelastic protons have, by definition, an energy lower than
elastic protons and therefore, a smaller velocity. They should then
appear on the right side of the elastic proton peak, corresponding to
a longer time-of-flight. In fact, due to the effect of the magnetic field and to the geometry of the \g0 collimators, inelastic protons have a shorter trajectory and they reach a given detector faster than the elastic ones. This is confirmed in the experimental data.

\subsection{Asymmetries}
\label{sec:6b}
The asymmetry is calculated using Eq. (\ref{eq97}).
with the following numerical values~\cite{mus94}:

${\sin}^{2}{\theta}_{_W}=0.23117$

${G}_{F}=1.16639\times 10^{-5}$

$\rvp=-0.0520$, $\rvn=-0.0143$, $\rvzero=0$.

\begin{figure}
\resizebox{0.55\textwidth}{!}{ 
\includegraphics{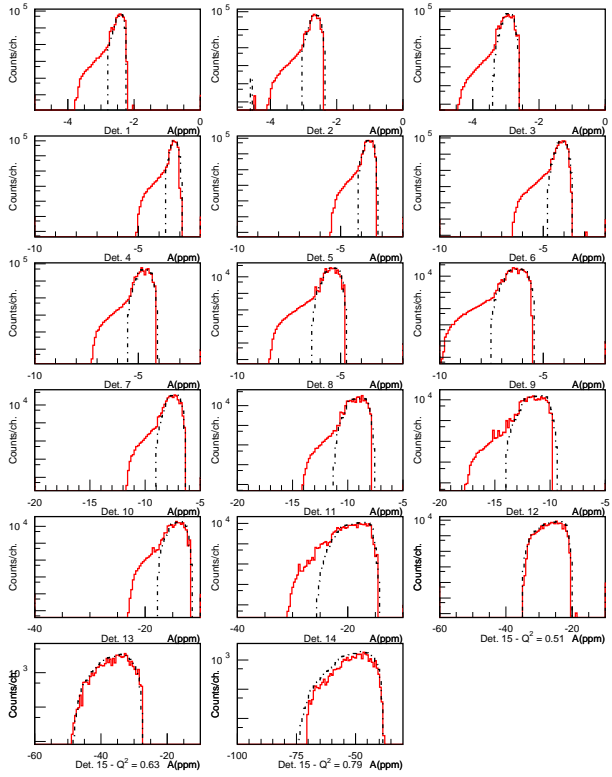}
}
\vspace{-1mm}
    \caption{Asymmetry distributions (in ppm), for each detector. The black dash-dotted line represent the elastic case, the red (grey) solid one the radiative case.}
   \label{asym_french}
\end{figure}

\noindent
$\ges$ and $\gms$ are parametrized with a dipole form according to \cite{mus94}.  A discussion of the latest electromagnetic form factors can be found in~\cite{arm05}.
The strangeness content parameters are from Hammer et al.~\cite{hmd96} with
$\mu _s = -0.24$ and $\rho _s = -2.93$. These values have been taken from a review paper by Kumar and Souder~\cite{ks00}. These parameters are used here only as an example of a strange asymmetry calculation. Electromagnetic radiative corrections are rather insensitive to the electroweak parameters.
Detailed calculations will be shown on the FR detectors spectra only. 
Fig.~\ref{asym_french} shows the asymmetry distribution  without (black line) and with RC (red/grey line). In the elastic case, the asymmetry reduces to the one calculated from the Born term only, and it can be compared directly to theoretical models. 

%%%%%%%%%%%%%%%%%%%%%%%%%%%%%%%%%%%%%%%%%%%%%%%%%%%%%%%%%%%%%%%%%%%%%%%%%%%%%%%%%

\begin{figure}
\resizebox{0.5\textwidth}{!}{ 
\includegraphics{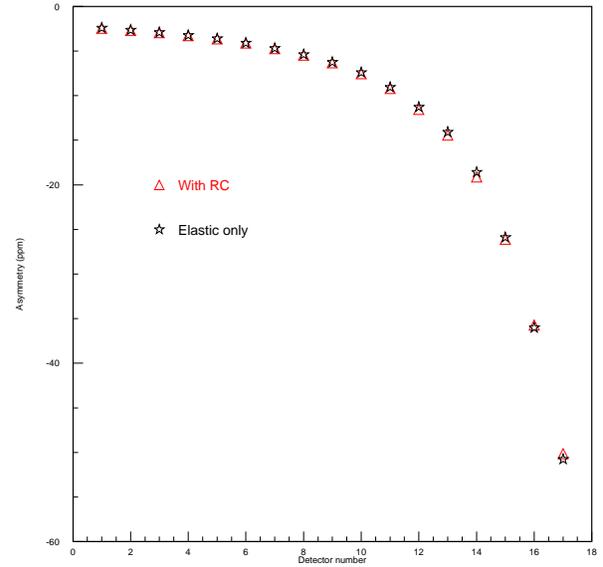}
}
\vspace{-1mm}
    \caption{Mean value of the asymmetry for each detector. The black stars represent the elastic case, the red (grey) triangles the radiative one. The ``french cuts'' on the tof distributions have been applied.}
   \label{asym_french_summary}
\end{figure}

\noindent
The mean asymmetry value is plotted in Fig.~\ref{asym_french_summary} for the FR detectors.
%%%%%%%%%%%%%%%%%%%%%%%%%%%%%%%%%%%%%%%%%%%%%%%%%%%%%%%%%%%%%%%%%%%%%%%%%%%%%%%%%%%%%
The effect of radiative corrections is to
increase the average asymmetry, following the increase in \q2. The
ratios between elastic and RC corrected asymmetries are given in the following Table \ref{table:rc-ratios}.

\begin{table} [ht]
\caption{\label{table:rc-ratios} Ratio of asymmetries $A_{el}/A_{RC}$ as a function of detector number where $A_{el}$ is the elastic (Born term) asymmetry and $A_{RC}$ is the asymmetry corrected for radiative emission.}
\begin{center}
\begin{tabular}{| c | c |}
\hline
{Detector \#} & {Ratio}\\
\hline
1&0.9971380\\    
2&0.9898130\\    
3&0.9912670\\  
4&0.9911590\\    
5& 0.9933250\\    
6& 0.9964800\\   
7& 0.9915390\\    
8& 0.9881630\\    
9& 0.9910010\\    
10& 0.9828710\\    
11& 0.9871740\\    
12& 0.9790010\\    
13& 0.9767610\\    
14& 0.9725560\\    
15/1 $Q^2=0.51$&  0.9922500\\    
15/2 $Q^2=0.63$&  1.008340\\    
15/3 $Q^2=0.79$&  1.012570\\      
\hline
\end{tabular}
\end{center}
\end{table}

\noindent
The `ideal' procedure to analyze the data would be to calculate an experimental asymmetry from the data after removing all background, leaving an experimental peak including radiative emission, and then to multiply the corresponding asymmetry by the ratio $A_{el}/A_{RC}$ given below. One problem is that, when removing background using a pure fitting procedure, one removes also part of the inelastic tail and therefore, by using the above procedure, one will overestimate the radiative corrections. A quantitative estimate of this effect is given below.

%%%%%%%%%%%%%%%%%%%%%%%%%%%%%%%%%%%%%%%%%%%%%%%%%%%%%%%%%%%%%%%%%%%%%%%%%%%%%%%
\noindent
 The asymmetry increase is of the
order of 0.5-1.0 \% for detectors 1-9, reaching 2.0 \% for detector 12\% and up to 3.0\% for detector 14. These ratios should be almost independent of the model chosen
and therefore valid for the no-strangeness value A$_0$. It is not clear if the dispersion between correction factors between 2 adjacent detectors (e.g. between Det. 8-9-10 or 10-11-12), which is of the order of 0.3 \%, is an indication of the present statistical/systematical errors or if it is a genuine effect due to differences in acceptance.

\noindent
\subsection{Uncertainty estimate}
\label{sec:7}
An error estimate is made based on the assumption that the elastic cuts have a 10\% uncertainty. Therefore the radiative corrections are calculated for cuts which are 5\% larger than the elastic cuts (by increasing the upper limit by 2.5\% and decreasing the lower limit by 2.5\%) and 5\% smaller than the elastic cuts (by decreasing the upper limit by 2.5\% and increasing the lower limit by 2.5\%). Then we take the ratio of these two quantities for each detector. This should represent an upper limit of the radiative correction uncertainties since the elastic cuts are known to better than 10\%.
The corresponding uncertainty would vary slowly from 0.1\% for Det. 1 to 0.5\% for Det. 13, 1\% for Det. 14 and between 0.0\% and 0.7\% for Det. 15, depending on the $Q^2$ cut. 
An alternative error estimate is obtained by making a global fit of the ratio $A_{el}/A_{RC}$ with a polynomial and assuming that the difference with the actual RC correction is due to systematics: in that case the uncertainty is globally estimated to be of the order of 0.1-0.3\% or 10\% of the actual correction depending on detector number. 

Another problem which has been investigated is the one of correction double-counting. If the background under the elastic peak is removed by a pure fitting procedure, it will also contain the RC tail contribution to the peak. Therefore the corresponding elastic asymmetry should not be corrected for RC effects. In order to estimate the sensitivity of the RC corrections, at the border of the elastic peak, we have calculated the RC corrections by adding or removing 1 ns from the elastic cuts. This effect has been estimated to be  about 2\% of the RC corrections which are themselves of the order of 2\%, so that double counting can be neglected at first order.

%%%%%%%%%%%%%%%%%%%%%%%%%%%%%%%%%%%%%%%%%%%%%%%%%%%%%%%%%%%%%%%%%%%%%%%
\section{Summary and conclusions}
\label{sec:8}
We have calculated the full electromagnetic radiative corrections for
elastic {\it ep} scattering in leptonic or hadronic variables; a performable
code has been constructed to extract the parity-violating asymmetry from
the experimental measured asymmetry. The comparison between the simulation
results and the data in the kinematic configuration of the PV-A4 experiment validates our procedure. 
Radiative corrections for the G0 parity-violating elastic scattering
experiment have been estimated by feeding our model calculations through
a Monte Carlo detector simulation. This code could also be used for the next
asymmetry measurement in the backward-angle configuration of G0.
\bigskip
%%%%%%%%%%%%%%%%%%%%%%%%%%%%%%%%%%%%%%%%%%%%%%%%%%%%%%%%%%%%%%%%%%%%%%%%%%%%%%%%%%%%%%%%%%%%%%%%%%%%%%

\begin{acknowledgement}

The authors are grateful to the PV-A4 and \g0 collaborations for their constructive remarks and support.
 
\end{acknowledgement}

\end{document}